\documentstyle[11pt,aaspp4]{article}

\cpright{AAS}{1997}
\slugcomment{To Appear in February 1997 A.J.}

\lefthead{Gizis}
\righthead{M-Subdwarfs}

\begin{document}

\title{M-Subdwarfs:  Spectroscopic Classification and the Metallicity Scale\footnote
{Observations were made partially at the 60-inch telescope at Palomar
Mountain which is jointly owned by the California Institute of Technology
and the Carnegie Institution of Washington}
 }

\author{John E. Gizis}
\affil{Palomar Observatory, 105-24, California Institute of Technology,
  Pasadena, California 91125, e-mail: jeg@astro.caltech.edu}



\begin{abstract}
We present a spectroscopic classification system for M-dwarfs and M-subdwarfs
based on quantitative measures of TiO and CaH features in the region  
$\lambda\lambda 6200 - 7400 \AA$.  Our sample of cool stars covers the
range from solar metallicity stars to the most extreme subdwarfs known.
Using synthetic spectra computed by Allard and Hauschildt (1995),
we derive metallicities for the stars.  Stars are classified 
as dwarfs (M V), subdwarfs (sdM), or extreme subdwarfs (esdM).
These classifications correspond to $[m/H] \approx 0.0$, -1.2, and
-2.0 respectively.    
Our metallicity scale agrees with theoretical HR diagrams
and HST globular cluster measurements.  We discuss some nearby
subdwarfs of particular interest in light of our metallicity scale.
\end{abstract}


\section{Introduction}

The vast majority of stars are M-dwarfs, main sequence stars whose spectra are
dominated by molecular absorption.  They have lifetimes much greater 
than the age of the universe which makes them an important fossil
record of Galactic history.  Their potential is largely unrealized 
because investigations of their properties have been hampered by 
the complex absorption spectra of diatomic and triatomic molecules.
In particular, theoretical model
atmospheres face serious difficulties; as a result,  
traditional methods of determining abundances from high resolution
spectra of weak atomic lines are inapplicable, requiring the use 
of other techniques. 

A fundamental astronomical tool is a classification system that
spans the range of observed properties, allowing 
an estimate of the effective temperature ($T_{eff}$), luminosity, and 
abundance ($[m/H]$) from a spectrum by comparison to standard
stars with known properties.  The system also should also 
allow identification of rare objects with unusual properties.  
The MKK dwarf spectral sequence (\cite{mkk}) extended
to M2 V and only included Population I objects.
As progressively cooler stars have been discovered
the classification system has been extended to M6.5 by Boeshaar (1976) 
and M9 (\cite{khm91}, hereafter KHM; see also Bessell 1991).  The
KHM system uses features in the wavelength ranges 
$\lambda\lambda 6950-7500 \AA$ and
$\lambda\lambda 8400-8950 \AA$ observed 
at $\sim 18 \AA$ resolution or better, and Henry { et al.}
(1994, hereafter HKS) have applied it to cool stars within 8 parsecs.
A slightly different approach was taken by Reid { et al.}
(1995, hereafter RHG) and Hawley { et al.} (1996, hereafter HGR)  
who used measurements of  the 7100 TiO bandhead at higher
resolution ($\sim 3 \AA$) to classify most of the known M-dwarfs within
25 parsecs.  The standards from KHM and HKS 
were used to place the RHG observations on the standard system.

The spectral classification system is thus now well-defined for
the near solar abundance M-dwarfs of the Galactic Disk.
The situation for low metallicity stars, traditionally called
subdwarfs, is much more confused.  Although many spectra of individual stars 
have been published, there is no consistent M-subdwarf
classification system.  Examples of metal-poor M-subdwarfs were
spectroscopically identified as long ago as Joy (1947), although ``later
investigators had difficulty in recognizing his criteria''
(\cite{o65}).  Mould and McElroy (1978) discussed ``old disk
subdwarfs'' which were less metal poor than other subdwarfs
on the basis of TiO and CaH indices.
Ake and Greenstein (1980), following a spectroscopic survey of high 
velocity stars, published spectra of four ``extreme subdwarf M stars''
which appeared to have ``extreme metal deficiency'' compared to  
the usually recognized M-subdwarfs.   Similar stars were 
identified spectroscopically in a search for nearby white dwarfs  (\cite{l79}),
a search for Population II halo stars (\cite{hcm84}), and a survey of
cool M-dwarfs (\cite{b82}), all targeted at faint, high-proper
motions stars.   Recently, trigonometric 
parallaxes of 17 extreme subdwarfs have been measured,
confirming their subluminosity (\cite{m92}, hereafter M92).
Thus objects selected by different criteria can be called subdwarfs by
different authors.  M92 also found that there is a gap between their 
extreme subdwarf sequence and the ``less extreme'' subdwarf and disk
sequence in the $M_V$ vs. V-I HR diagram -- their preliminary 
interpretation was that this gap represented a real lack of 
``intermediate'' metallicity stars.  In any case, the physical properties
of the various subdwarf types 
remain in doubt; for example, M92 argue that their
extreme subdwarfs have $[m/H] \sim -1.7$ whereas Eggen (1996) 
suggests they have $[m/H] \sim -2.5 $ to $-3.5$. 

Here we present a self-consistent set of spectroscopic observations 
of cool metal-poor stars drawn from a variety of sources.   
Selection of our subdwarf candidates and the data reduction
procedures are discussed in Section~\ref{secobs}.
In Section~\ref{secbands}, we discuss empirical molecular 
bandstrengths and suggest a two-dimensional classification system.  
In Section~\ref{secmodel},
we compare our spectra to model atmospheres and deduce metallicities.
In Section~\ref{sechr}, we discuss the use of color-color and HR diagrams.
Some notable individual stars are discussed in Section~\ref{secnotes}.
The results are summarized in Section~\ref{conclusions}.

\section{Observations\label{secobs}}

\subsection{Sample Selection}

Identification of a metal-poor or halo population can be difficult
and ambiguous.
It is well known from studies of hotter (F, G, and K) stars
that there is no one-to-one correspondence between kinematics
and metallicity because there is overlap between the properties
of the disk and halo components (e.g., Mihalas and Binney 1981).
Even among local stars with $v_{tan} > 100 {\rm ~km~s}^{-1}$, 
the halo (Population II) is outnumbered approximately
ten to one (\cite{s75}) by high velocity, slightly metal-poor
disk stars (the Intermediate Population II, reviewed by \cite{m93}). 
Despite the name Intermediate Population II (IPII), these stars have
a mean metallicity ($[m/H] \sim -0.6$) close to that of the disk
($[m/H] \sim 0.0$), although the exact abundance distribution of
the IPII is difficult to determine (\cite{cll89}).  
In contrast, a conservatively selected field halo sample  (\cite{lrcl88})   
(restricted to stars that have $v_{tan} > 220 {\rm ~km~s}^{-1}$ or 
$V<-220 {\rm ~km~s}^{-1}$ 
\footnote{We use the standard notation of (U,V,W)
for the space velocity components; note that 
U is positive towards the Galactic Center ($l=0$,$b=0$).})
has a distribution in $[m/H]$ that peaks at -1.7 with FWHM $\sim 1.2$ dex, and 
includes tails that extend down to very low abundances
($8 \%$ of the stars have $[m/H]<-2.5$) and up to IPII-like abundances
($9 \%$ have $[m/H] > -1.0$).  
We therefore
{\it a priori} expect that kinematically selected samples with
loose selection criteria (e.g., $v_{tan} > 100 {\rm ~km~s}^{-1}$) 
will include many IPII stars that have
$[m/H] > -1$, and we also expect to see a large range of metallicities
even among ``true'' halo stars.

We have chosen our objects from a number of sources; however, virtually
all have been identified in proper motion surveys and appear in the
Luyten LHS catalog (\cite{lhs}) and most are in the 
the Lowell Proper Motion Survey (\cite{giclas}).
These surveys give proper motions, photographic
magnitudes, and low-precision color indices which are by themselves not
adequate to isolate a halo or metal-poor sample.
However, many late-type candidate subdwarfs have been identified in 
followup surveys.  Since we use a variety of sources, 
selection criteria are
ill-defined but typically depend on the star's velocity, color, or
spectral features.   We particularly favor stars with measured 
trigonometric parallaxes.  These effects are not important for this work
since we are not setting out to measure statistical quantities. 
Of particular note are the Schmidt (1975)
complete sample of stars with $\mu > 1.295 \arcsec$ yr$^{-1}$ and
$m_{pg} < 15.95$,
the Greenstein (1989) sample of cool halo stars, and the M92 
and Ruiz and Anguita (1993, hereafter RA)
samples of CCD parallax stars.  Other objects have been chosen 
from lists of unusual stars in photometric followups
(\cite{df89,df92,r82,l92})   Finally, it should noted that 
the Giclas catalog only goes to $m_{pg} \sim 17$ and is
incomplete for $m_{pg} \ge 16$.  Schmidt (1975) showed that
the apparent lack of $M_{pg}>13.5$ ($M_V \sim 12$) is due to this
apparent magnitude limit.  The Luyten searches
reach $m_R \sim 20$ but are incomplete both for $m_R > 18$
and $\mu > 2.5 \arcsec$/yr (\cite{d86}).  The latter limit implies
that stars with typical halo tangential velocities of 
$220$ and $300 {\rm ~km~s}^{-1}$   
would not have been detected within 18 and 25 parsecs respectively.

In addition to the candidate subdwarfs selected above, 
we have utilized the RHG observations of 1700 M-dwarfs from 
the preliminary Third Catalog of Nearby Stars (\cite{gj91}, CNS3)
as a reference solar abundance sample.  
We note particular use of two subsets from this sample.
First, we use the stars within eight parsecs, selected to be 
either single or well separated from their companions, which outline the
detailed structure in the HR diagram (e.g., \cite{gr96}).
Distances were adopted from RHG, but the 
stars were required to have accurate trigonometric parallaxes
with Lutz-Kelker (1973) corrections less than 0.1 magnitudes. 
Second, we also use the nearby stars with $v_{tan} > 100 {\rm ~km~s}^{-1}$.

\subsection{Spectroscopic Observations and Data Reduction}

Spectra were obtained at the Palomar 60 in. telescope, the Hale
200 in. telescope, and the Las Campanas Du Pont 100 in. telescope.  
We used the G-mode of the Palomar 60 in. spectrograph
(\cite{m85}), a 1 arcsecond slit, and a 600 l/mm grating blazed at
$6500 \AA$, yielding $1.5~\AA~{\rm pix}^{-1}$.  Candidate subdwarfs were
observed in May 1994, September 1994, and January 1996 while
some CNS3 stars were observed with the same setup in
1993 and 1994 (RHG). At the Hale 200 in. telescope we used the
double spectrograph.   In August 1995 with the blue camera was set to 
observe $6000-6900 \AA$ and the red camera was set to  $6700-8000 \AA$
using $600$ l/mm gratings blazed at $4000 \AA$ and $10000 \AA$ respectively. 
In October 1995, a new red camera was installed in the
double spectrograph, which we used in all subsequent runs to observe the region
$\lambda 6000-7400 \AA$ at $1.4~\AA~{\rm pix}^{-1}$. 
The 100 in. telescope observations used the
modular spectrograph with a 1200 line grating blazed at $7500 \AA$.
The resolution at all telescopes was $3-4 \AA$.  
At all telescopes, neon and argon arcs were
taken after each observation in order to eliminate the effects of
instrument flexure.  The data were extracted, sky subtracted,
and wavelength calibrated using the FIGARO
package.   Flux standards (\cite{go83} and \cite{bs84}) were used to
set the data on an $F_\nu$ scale.  Since observing conditions were
usually non-photometric with seeing worse than 1 arcsecond, the
fluxes are not absolute.  

Radial velocities were measured using the FIGARO cross correlation
task SCROSS and corrected to a heliocentric frame using 
VHELIO.  M-dwarf radial velocity standards were drawn from
Marcy and Benitz (1986).  For the Hale 200 in. telescope observations,
the extreme subdwarf LHS 1174 was also used as a cross correlation
standard.  A radial velocity of $-112 \pm 2 {\rm ~km~s}^{-1}$ was adopted 
based upon a Keck HIRES spectrum (Reid, personal communication).
Using either the extreme subdwarf or disk M-dwarfs
did not make a significant change in the derived velocities,  
which have an uncertainty of $\pm \sim 20 {\rm ~km~s}^{-1}$.  
We have 14 stars 
in common with the more precise measurements of Dawson and 
De Robertis (1988, 1989) -- we find a mean difference of 
$10 {\rm ~km~s}^{-1}$ and a standard deviation of $26 {\rm ~km~s}^{-1}$.  
The three stars with differences of
more than $30 {\rm ~km~s}^{-1}$ are LHS 161, 205a, and 479.

\section{Bandstrengths and a Subdwarf Spectroscopic Sequence\label{secbands}}

Molecular features are typically very broad (tens or hundreds of
Angstroms) and often are asymmetric.  In M-dwarfs, there are few points
that are relatively clear of absorption, but model atmospheres
(Allard and Hauschildt 1995) show that even these are depressed
from the "true" continuum.  Since absolute measurements of the strength of
features are therefore impossible, we instead measure
molecular features with bandstrength indices defined as 
$$R_{ind} = {{F_{W}}\over{F_{cont}}}$$ where pseudo-continuum ($F_{cont}$)
region and the feature ($F_{W}$) wavelength limits are taken from RHG
and listed here in Table~\ref{index}.
In the case of the CaH1 index, two sidebands are used to estimate
the pseudo-continuum flux ($S_1$ and $S_2$).  
As in RHG, the spectral indices have  
an accuracy of $\pm 0.02-0.04$.  Note that our experience with these
indices show that all except CaH2, which uses widely separated
pseudo-continuum and feature regions, are insensitive to the 
flux calibration errors.  CaH2 shows systematic offsets of up to $0.03$
from observing run to observing run.  The measured indices for our stars 
are given in Table~\ref{spectra}.

The KHM M-dwarf spectral types provide an excellent shorthand description
of the spectral properties of Galactic Disk stars -- 
we find that over the range
K7 - M6 the relations
\begin{equation}
\label{tiofit}
Sp = -9.64 \times \rm{TiO5} + 7.76
\end{equation}
\begin{equation}
\label{cah2fit}
Sp = 7.91\times \rm{CaH2}^2 - 20.63 \times \rm{CaH2} + 10.71
\end{equation}
\begin{equation}
\label{cah3fit}
Sp = -18.00 \times \rm{CaH3} + 15.80
\end{equation}
are accurate to $\pm 0.5$ subclass, where K5 and K7 are $Sp = -2$ and -1
respectively.  As described by RHG, the CaH1 index saturates at $\sim M3$,
beyond which it does not show much temperature dependence.
Although TiO5 was used by RHG to classify the stars, the two
CaH relations are equally good spectral type indicators for near
solar metallicity stars. 

\begin{figure}
\plotone{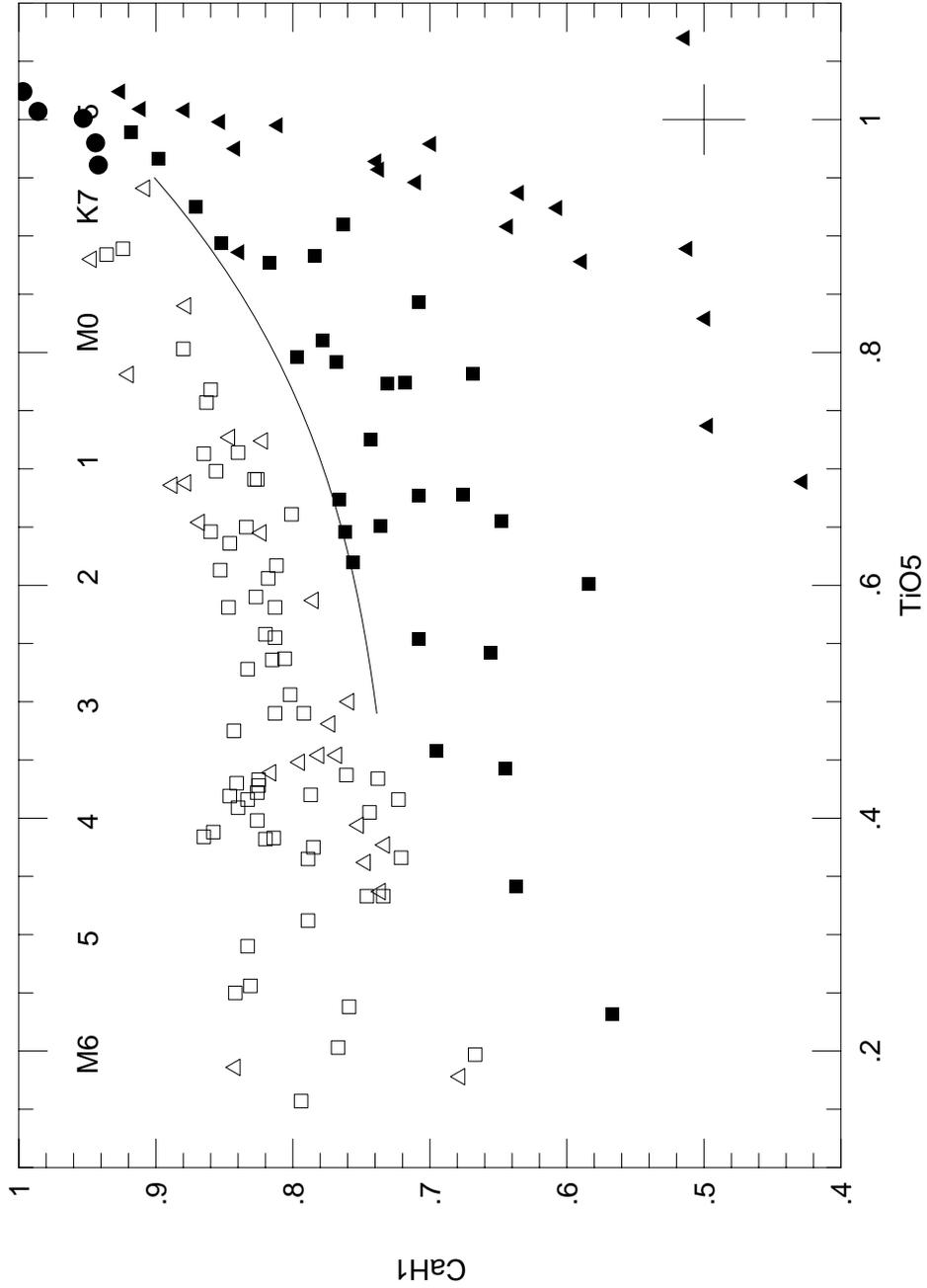}
\caption{The CaH indices as a function of TiO5.  Stars classified 
as sdM appear as filled squares, esdM as filled triangles, sdK
as filled circles, and 
spectroscopic non-subdwarfs as open triangles.  In the CaH1 diagram
we also plot the single stars within eight parsecs as open squares.
All other figures use the same symbols.  Representative error bars
($\pm 0.03$) appear in the lower right.  
\label{cahtiofig}}
\end{figure}

Figure~\ref{cahtiofig} shows the three CaH indices plotted against TiO5 for our
candidate metal-poor stars and the eight parsec disk sample.
\footnote{The relations defined by all of the M-dwarfs in the northern CNS3 
are plotted in  RHG's Figure 4.}
Fifty stars lie significantly below the mean CaH vs TiO5 relations,
a significant difference compared with only 5 of 1685 stars in RHG.
There appears to be a fairly well defined sequence to the 
lower right of each figure -- in Figure~\ref{cahtiofig}b a high-order 
polynomial illustrates the cutoff adopted.  We will call the stars below
this cutoff the extreme
M-subdwarfs.  Note however that the region between the extreme M-subdwarf
sequence and the disk M-dwarfs is well represented in our sample.       

The extreme subdwarf sequence, with the notable exception of LHS 453, 
has very weak but detectable TiO 
absorption at the resolution of our observations.  
However, even the coolest star, LHS 1742a ($M_V = 14.43$), 
has only the TiO strength of a disk
M1 star -- the rest of the sequence is equivalent to K7 or K5 stars
according to equation~\ref{tiofit}.  
Indeed, LHS 1174 was classified ``sdK7-M1'' 
by Liebert (1991, personal communication reported in M92).
A TiO classification thus compresses the entire sequence of M-subdwarfs
into only three subclasses even though Figure~\ref{cahtiofig} shows there is
as much or more variation in the CaH strengths 
as TiO variation in the disk M-dwarfs.  
Use of equations~\ref{cah2fit} or \ref{cah3fit} gives spectral types
of M5.5 (CaH2) or M8 (CaH3, extrapolated) for LHS 1742a.  

In many circumstances a simplified
shorthand notation may be more useful than measurements of 
specific molecular features, even though Figure~\ref{cahtiofig} 
shows that there is a full two-dimensional continuum in the 
properties of M-subdwarfs.  Using the TiO
and CaH indices as a guide, a sequence of low abundance M spectral standards
can be defined.  While this is the first subdwarf standard sequence
to be defined, historical precedent 
supports using the prefix ``sd''before the M spectral class.
(Luminosity class VI should not be used for subdwarfs according
to \cite{jj87}.)   
In the past, LHS 64 has been classified sdM1 (\cite{j47}), very
similar to the ``sdK7-M1'' classification for LHS 1174.
Unfortunately the use of ``sd'' alone throws away
much information -- although both types of subdwarfs 
above are significantly different
from Population I stars, they also differ at least as much from each other,
as seen in Table~\ref{spectra} and Figure~\ref{cahtiofig}.   
Second, as discussed above, use of TiO alone would artificially 
suppress much of the variance among the extreme subdwarfs.

We therefore propose the following classification system, which consists
of both a prefix (M, sdM, esdM) and a numerical subclass.  The classification
system works as follows.  First, determine if the star if a subdwarf
of some kind using CaH1.  Second, use CaH2 to determine if the
star is a sdM or esdM. Finally, assign a numerical subclass.   
Each step is described in detail below.  

In the first step, stars that show significantly stronger CaH absorption 
for their TiO5 strength compared with disk dwarfs are
designed subdwarfs with the prefix  ``sdM'' or ``esdM''.  Quantitatively, we
define significant as lying 0.07 below the mean Population I CaH1-TiO5
relation of the RHG stars.  If the other indices are unavailable, 0.06 below 
the RHG CaH2-TiO5 or CaH3-TiO5 relations is equivalent.  For the coolest
stars ($TiO5<0.49$), CaH2 or CaH3 must be used since the disk CaH1
relation is ``saturated.''  The equations for these cutoffs are:
\begin{equation}
\label{cah1out}
\rm{CaH1} < 0.695 \times \rm{TiO5}^3 -0.818 \times \rm{TiO5}^2 + 0.413 \times \rm{TiO5} +0.651
\end{equation}
\begin{equation}
\label{cah2out}
\rm{CaH2} < 0.968 \times \rm{TiO5}^3 -1.358 \times \rm{TiO5}^2 + 1.315 \times \rm{TiO5} -0.033
\end{equation}
\begin{equation}
\label{cah3out}
\rm{CaH3} < 0.639 \times \rm{TiO5}^3 -1.199 \times \rm{TiO5}^2 + 1.161 \times \rm{TiO5} +0.307
\end{equation}
Stars that fulfill Equations~\ref{cah1out},~\ref{cah2out}, or~\ref{cah3out}
are either sdM or esdM.  Equation~\ref{cah1out} is plotted in 
Figure~\ref{cahtiofig}a.    
In the second step, the extreme subdwarfs are identified.
A high order polynomial, shown in Figure~\ref{cahtiofig}b, was used
to interpolate between 
arbitrarily chosen points to define a cutoff in CaH2 vs. TiO5.  
The coordinates of the points, expressed as (TiO5,CaH2), are
(0.0,-0.1),(0.433,0.234),(0.600,0.303),(0.800,0.456),(0.905,0.626),
(0.977,0.788), and (1.020,0.914).  The sdM lie above these
points and the esdM lie below.  
The separation between the subdwarfs and extreme subdwarfs is natural 
(at least for this sample)
for the M stars, but at the hotter end the weakness of the 
features makes the sequences merge.  At the cool end,
the relation is not defined beyond ${\rm TiO5} \sim 0.6$,
but the points given prevent the interpolating polynomial
from inconveniently crossing the disk sequence.

Finally, the numerical subclass is determined.  
The definition of the numerical 
subclasses is problematic.  Larger numbers correspond to lower
temperatures, but since effective temperatures are not well determined
for cool stars it is impossible to ensure that sdM subclasses correspond
to the same temperatures as M V stars or esdM stars.  Some authors
(e.g., \cite{hs93}) have used pseudocontinuum indices to assign types for
late-type disk M-dwarfs, but we find that the pseudocontinuum points available
in our spectral range show more scatter than the bandstrength indices
we are using.  We adopt
Equation~\ref{cah2fit} as the definition of numerical subclasses
for the sdM and esdM because this feature shows the same range 
($0.2 \lesssim \rm{CaH2} \lesssim 1.0$) for all metallicities.  Use of 
the disk relation for
the CaH3 feature (Equation~\ref{cah3fit}) or the TiO5 feature
(Equation~\ref{tiofit}) leads to very large and very small ranges in
the spectral type for the esdM.  The subclass, however, is 
relatively uncertain when using only the CaH2 feature 
since the spectral type determination then depends 
upon only a single feature.  We have therefore determined the 
following linear fits:
\begin{equation}
\label{cah3fitforsdm}
Sp_{\rm{sdM}} = -16.02 \times \rm{CaH3} + 13.78
\end{equation}
\begin{equation}
\label{cah3fitforesdm}
Sp_{\rm{esdM}} = -13.47 \times \rm{CaH3} + 11.50
\end{equation}
The numerical subclasses listed in Table~\ref{spectra} are the average 
of the CaH2 relation (Equation~\ref{cah2fit}) and the appropriate
CaH3 relation (Equation~\ref{cah3fit},~\ref{cah3fitforsdm}, or
~\ref{cah3fitforesdm}).  We continue to classify 
stars as ``sd'' or ``esd'' as early as K7.  For earlier stars,
we assign classes of only sdK (or sdG for stars with strong $H \alpha$ 
absorption).  The cooler sdK can be 
distinguished in Figure~\ref{cahtiofig} where they have stronger
CaH1 relative to the K stars presented in RHG.  
These features are quite weak so the measurement errors are
more important.  The classifications of the K stars
are therefore not as certain and should be regarded with caution --
they are presented here to show the continuity with the
M-subdwarfs.  The new classifications are listed
in Table~\ref{spectra}.  Figures~\ref{figsdm} and \ref{figesdm} show 
sequences of spectra
for the subdwarfs and extreme subdwarfs respectively.  
  
The cool ends of the sequences are not yet well-defined, since there
are as yet observations of only a few very cool spectroscopic subdwarfs.   
LHS 407 has been classified as sdM5.0.  
Although in Figure~\ref{cahtiofig} it is well to the left
of the esdM, it lies with the extreme subdwarfs in Figure~\ref{ratio}
and the HR diagram (Figure~\ref{hr}).  It is easily distinguished from
LHS 1742a (esdM5.5), which is at present the coolest extreme subdwarf known.
LHS 3061 is similar enough to LHS 1742a 
to also be esdM5.5, but it is slightly hotter 
according to its spectral indices and M92 V-I color ($\Delta_{V-I}=0.13$).
LHS 3409, observed by RHG, is apparently more metal-rich than LHS 407,
yet its indices are significantly different from the Population I,
or even most stars with velocities typical of the IPII.  We have therefore
called it sdM4.5. LHS 3480 is a similar star.    
A considerably redder sdM is the star LHS 377
(shown by M92 to have $M_V = 15.66$).  Its TiO5 index is equivalent to 
an M5.5 V star, but CaH2 gives the classification sdM7 (M92 report that
Boeshaar and Liebert call the star sdM5 in a private communication).    
Finally, we have also observed TVLM 832-42500 which is subluminous
in the I-K vs $M_K$ diagram, which suggests it has
$[m/H] \sim -0.5$ to $-1.0$ (\cite{trgm95}).  It is at spectral type 
$\sim $ M7 V  where the indices break down; however, compared to the
RHG stars of similar TiO5 strength 
it has slightly stronger CaH2 and CaH3, suggesting it is indeed
slightly more metal poor, or at least spectroscopically different.
Its TiO5 and CaH2 absorption is slightly stronger than LHS 377
but the CaH3 absorption is much weaker.  It thus appears to be
more metal rich than LHS 377.  A proper
understanding of the distinction between cool sdM vs. esdM will require 
identification of more very cool stars from the LHS catalog.  However, since we
determine the M spectral subclass by the quantitative CaH2 index,
the classification of M4.5 and M5.5 for the stars above is well-defined
in our system.  

The relative strengths of bandstrength indices contains more
information than just a simple classification.  HGR
show that there are systematic differences in the TiO indices
of dMe and dM stars.  In Figure~\ref{ratio}, we plot the ratio CaH3/CaH2
as a function of CaH2 strength for all the RHG stars
and for the sdM and esdM.  The sdM tend to lie below the Population I 
sequence.  The extreme M-subdwarfs lie systematically well 
below the sequence.  The trend is that the first band of CaH (CaH3)
is stronger relative to the entire bandhead (CaH2) -- similar to the
behavior seen in TiO for late type M-dwarfs.  Since the
deeper band must reflect conditions higher in the atmosphere,
the difference might be caused by a slight surface gravity difference
and/or a lower temperature 
in the CaH3 formation region compared to sdM and dM.  Detailed
modeling is necessary to understand this difference.

\section{Model Atmosphere Fitting\label{secmodel}}

Interpretation of the observed spectra in terms of 
the physical parameters effective temperature ($T_{eff}$)
and metallicity ($[m/H]$) requires the use of model atmospheres.  
In principle, surface gravity is
also a relevant parameter; however, both theory and observations
indicate dwarf stars have $\log g \sim 5.0 \pm 0.2$ (Leggett { et al.}
1996) and we therefore restrict analysis to 
$\log g = 5.0$ models.  We use the Extended Model grid
computed by Allard and Hauschildt (1995, hereafter AH) which
assumes local thermodynamic equilibrium (LTE).

The theoretical uncertainties and limitations of the atmosphere
models have been extensively discussed by AH.  Although there are
uncertainties in the appropriate treatment of the physics
such as non-LTE effects,
the treatment of molecular opacities (usually the Just Overlapping
Line Approximation is used in these models), and the treatment of convection,
the primary limitation is simply the
lack of both laboratory data and theoretical computations of the
electronic band oscillator strengths.  In particular, no data
exist for the CaH molecule which is extremely important in our
spectral region. 
In Figure~\ref{fitpopi}, we show a synthetic spectrum based upon a 
solar metallicity model and a representative disk star, 
the KHM M 4.0 V  standard Gl 402.     
(The treatment of the TiO molecule in solar metallicity models has recently
been improved by Allard and Hauschildt in the ``NextGen'' model).
The spectra are in good general agreement, although the specific strengths
of the molecular features are incorrect, as expected.  In particular,
the CaH feature at $\sim 6950 \AA$ is too strong in all models
compared to the observed spectra.  However, AH show that
their models capture the general overall behavior of metal-poor stars --
with decreasing metallicity the hydride bands and atomic lines both
increase in strength -- the latter due to an increase in gas pressure.

Figures~\ref{figsdm} and \ref{figesdm} 
clearly show the expected increase in 
the atomic line strength with increasing hydride strength.  
Until accurate data or calculations are available for all the important 
molecules, but especially the hydride bands, any comparison between
observations and theory will remain uncertain.  
Since such information is not likely to be available for many years,
the uncertainties of this study will not reduced in the near future.  
Other authors have compared the AH models to 
cool solar abundance and metal-poor dwarfs
using broad band colors and low-resolution
optical and infrared spectra (\cite{dlhg94}; \cite{l96}) or higher resolution
infrared spectra (\cite{jones}).  Our study is most sensitive to
variations in CaH and TiO which dominate our more limited wavelength
region.  It should also
be noted that the metal poor models are computed assuming 
scaled solar abundances.   
If $[O/C]$ is enhanced in the Population II M-subdwarfs,
as it is in hotter Population II stars, it
will affect the strength of TiO and $\rm{H}_{2}\rm{O}$ absorption
(\cite{m76}).

Due to the uncertainties in the molecular data, the AH models cannot
be used to predict the bandstrengths measured in Section~\ref{secbands}, 
nor are they
able to reproduce the CaH3/CaH2 relation in Figure~\ref{ratio}.
Instead, we compare the spectral region ($\lambda\lambda 6200-7300$)
by minimizing the least squares difference ($\chi^2$) 
between the observed spectrum
and the models.  We first shift the observations to rest velocity
and vacuum wavelengths, convert
to $F_{\lambda}$, and rebin to $2 \AA$ steps.  The models are convolved
with a $\sigma = 2\AA$ Gaussian to approximate the instrumental 
resolution (in practice, this has negligible effects since the synthetic
spectra often show $50 \AA$ wide ``features'' due to the 
computation technique).
The extended models have been computed for steps of 0.5 dex in 
metallicity between $[m/H] = 0.0$ and $[m/H] = -4.0$.  We consider
$100 K$ steps for the temperature range $2500 K \le T_{eff} \le 4000 K$ (higher
temperatures are not available and lower temperatures are irrelevant
to the stars under consideration here).    For temperatures
that were not calculated ($T_{eff}$ =  2600, 3100, 3400, 3600, 3800,
 and 3900 K),
we approximate the synthetic spectra by linearly interpolating between the two
nearest temperatures.   

The best fit model parameters for the subdwarf candidates 
are listed in Table~\ref{fit}.  We have also fit the eight parsec
sample of single disk M-dwarfs; below 3800 K, the spectra are best matched
by solar metallicity models, as expected.  Above 3800 K, the fits are
poor because the molecular features are too weak to constrain the
fits well. We therefore report only those fits with $T_{eff} < 3800$ K. 
The uncertainties from the fitting procedure are the grid spacing
of $\pm 0.5$ in $[m/H]$ and $\pm 100$ K in $T_{eff}$.  
The systematic uncertainties
are unknown, but are dominated by the errors in the synthetic spectra
discussed above.  We note that
different temperature scales for the disk M-dwarfs have historically
had systematic differences of up to 300 K (e.g., \cite{l96}) and our
fits are subject to all the same uncertainties.  Compared to 
Leggett { et al.} (1996),  our fits give somewhat higher 
temperatures ($\Delta T_{eff} = 150$K for LHS 377, 
$\Delta T_{eff} = 400$ K for LHS 57).  

The extreme subdwarfs have best fits at $[m/H] \sim -2$.  
Figure~\ref{fitlhs205a} shows one representative fit..  The coolest 
extreme subdwarf,
LHS 1742a, has an effective temperature of 3300 K.  A difference
of one extreme subdwarf spectral subclass derived from CaH2 corresponds to an 
approximately 100 K difference in temperature.    
Some authors (e.g., \cite{e96}) have argued that the the extreme
subdwarfs are likely to have $[m/H] = -2.5$ to $-3.0$, whereas our
least-squares fitting of model atmospheres gives metallicities a
factor of 10 higher.   
The model atmospheres show that the most obvious qualitative difference is that
$\sim 7100 \AA$ TiO features for the -2.5 and -3.0 models should be completely
absent.   The observed spectra, however, show that although the feature is
weak it does exist in our extreme subdwarfs (except LHS 453), supporting the
model atmosphere fits.  LHS 453 completely lacks the TiO5 feature
(the TiO feature at $6200 \AA$ is also missing) and therefore is more
metal poor than the other esdM.  The best fit is $[m/H] = -2.5$.  
For higher metallicities
($[m/H] = -1.0$) the TiO feature remains strong which is 
qualitatively seen in the stars classified sdM.   Indeed, the sdM have best
$[m/H]$ fits of -1.0 to -1.5.  For the stars that were not classified
subdwarfs, the model fits give $[m/H]$ of 0.0 to -1.0.  

In summary, the spectroscopic classification system 
of Section~\ref{secbands} can be used to derive 
metallicities based upon our fits to AH synthetic spectra.  
Stars classified as
extreme M-subdwarfs (esdM) have $[m/H]$ of $-2.0 \pm -0.5$,
whereas sdM have a metallicity of $-1.2 \pm 0.3$.  Stars that do not
show significant differences from the RHG disk sample
in their CaH indices have $[m/H] > -1.0$.  
When model atmospheres are able to accurately reproduce the 
CaH and TiO bandheads more accurate metallicities should be possible.    

\section{HR diagrams\label{sechr}}

HR diagrams are an important tool for the study of cool stars.  
Observational HR diagrams are a critical test of  
stellar structure theory, provided that the transformation
between theoretical temperatures and luminosities and
observed quantities (colors and absolute magnitudes)
can be made.  In practice there has been
disagreement over the Population I temperature scale (\cite{l96})
and relatively little discussion of the temperature of Population II
subdwarfs.   The HR diagram can also be
used to predict absolute magnitude from observed color and thus is used
in interpretation of star count data and derivation of the
luminosity function.   Finally,   
derivation of the mass function 
from the luminosity function of metal-poor stars 
depends upon the quality of the 
theoretical mass-luminosity relations and evolutionary tracks since
there are no cool Population II binaries with mass determinations.  

Table~\ref{photo} lists optical photometry and
trigonometric parallaxes from the literature 
for the stars in our sample.  We have adopted the Cousins R and I system
and where necessary have used the color transformations
compiled by Leggett (1992).  Because the photometry is from
heterogeneous sources of varying quality, in some cases the
colors may be incorrect by as much as 0.1 magnitudes 
but the typical accuracy should be about $5 \%$ (\cite{l92}).
In addition, the transformations have been derived for 
near solar metallicity stars and may be inaccurate for subdwarfs.
Absolute parallaxes were taken from the preliminary Yale
Parallax Catalog (\cite{va91}) when available.  In the case of
the CCD relative parallaxes reported by RA,
we added a correction of 0.5 milliarcseconds to the parallaxes to 
transform them to absolute measurements.  This 
approximately matches the correction applied by M92 for
similar magnitude stars and the uncertainty in this correction is
much less than the uncertainties in the relative parallaxes.
Although the measured parallax is the best estimator of the
distance to any particular star, there is a statistical bias 
to underestimate the distance and hence underestimate the
the mean luminosity of the sample stars.  Lutz and Kelker (1973) 
derived statistical corrections to the observed absolute magnitudes
as a function of $\sigma /  \pi$ and showed that only parallaxes with
$\sigma / \pi < 0.2$ are useful.  However, selecting the stars 
by proper motion reduces the statistical correction (\cite{h79}).
Since the selection of our sample is not well-defined, 
we do not apply any statistical correction to the absolute magnitudes,
but we list the Lutz-Kelker corrections (derived from the 
approximation given by \cite{h79} and defined as 
$M_{true} = M_{obs} + \Delta_{LK}$) as a guide to likely biases
in the sample.  
The V-I, $M_{V}$ diagram is shown in Figure~\ref{hr}.  The eight parsec
single stars are plotted with Lutz-Kelker corrections applied, but
the sample is restricted to stars with corrections less than 0.1 magnitude.
We also plot the B-V version of this diagram in Figure~\ref{hrbv}, 
including the bluer K subdwarfs in the sample and
the disk G and K dwarfs within eight parsecs from the CNS3.
It is well known that B-V colors are poorly suited for
estimating absolute magnitudes for the disk M-dwarfs (in the disk
sequence $M_V$ changes by 4 magnitudes in the range $1.4 < \rm{B-V} < 1.6$.)
In our diagram, the sdM and esdM sequences cross the disk sequence at
B-V $\sim 1.5$, and the reddest sdM and esdM actually lie above
the disk sequence for B-V $> 1.65$. 

Optical and near-infrared color-color diagrams for M-dwarfs
have recently been extensively discussed by Leggett (1992).  
In Figure~\ref{vi-bv} we show the V-I vs. B-V diagram for the objects
we have classified.   The well-known tendency for metal-poor M-subdwarfs
to have redder B-V at a given V-I (e.g., \cite{mm78}; \cite{dlhg94}) 
is clearly evident, for $V-I > 1.7$.  The 
offset appears to be strongly related to metallicity, as the esdM
are approximately 0.3 magnitudes redder than the disk stars 
in B-V with our sdM lying in between the two sequences.
A V-I vs. B-V diagram is therefore useful in identifying 
red subdwarfs and extreme subdwarfs.  However, giants also lie 
redwards in B-V at a given V-I (\cite{w77}).
Indeed, Reid (1982) suggested on the basis of optical (BVRI) colors  
that Sm 183 might be a subdwarf similar to Kapteyn's star, 
but also noted the possibility that it could be a giant.
In our spectrum, the CaH indices are weak compared to 
TiO, implying that it is a low surface gravity giant.

\subsection{Tests of the Spectroscopic Metallicity Scale}

Theoretical structure predictions and HST globular cluster 
observations provide checks on our spectroscopic metallicity scale.
We compare with broadband colors and magnitudes 
for $[m/H] = 0.0$, -0.5, and -1.5 computed with 
stellar interior models that use  
the Allard and Hauschildt model atmospheres as outer boundary conditions
(\cite{bcah95}).  The tracks are shown in Figure~\ref{hrtheory}.  
Baraffe et al. (1996) did not compute $[m/H] = -1.0$ models, but they
did note that for 
$[m/H]= -2.0$ the models are shifted blueward by $\sim 0.2$ magnitudes
with respect to the -1.5 track.  Also shown
is an even more recent set of solar metallicity models computed
with the latest Allard and Hauschildt NextGen models (\cite{bc96}).  
The results dramatically illustrate the importance of boundary conditions  --
at $0.3 M_{\odot}$ ($M_V = 11.31$, V-I=2.19, $T_{eff}=3403$), 
the newer model is 201 K hotter, 1.21 magnitudes
brighter in $M_V$, and 0.63 magnitudes bluer in V-I. 
As a result, the newer models fail to match
observations (suggesting that they would be a poor choice to
use for the $M_V$-Mass relation).
In our HR diagram, the earlier set of computations
give better agreement with the Population I sequence, although none
of them convincingly duplicate the structure seen.  As it stands, 
these models support the identification of the esdM 
with $[m/H] \sim -1.5$ to $-2.0$ and the sdM as
more metal poor than -0.5.  Given the change of the theoretical
models in less than one year, effective temperatures, metallicities,
and masses derived from evolutionary models are clearly
very uncertain -- but the use of Allard and Hauschildt's extended
model grid leads to consistent results for both spectroscopic
and HR diagram metallicities.  

HST photometry of a stellar cluster yields both a luminosity function
and a color-magnitude diagram.   Santiago { et al.} (1996, hereafter SEG)
provide piecewise linear fits to the HR diagram of the 
clusters M15 ($[m/H] = -2.26$),
$\omega$ Cen (-1.6), 47 Tuc (-0.6), NGC 2420 (-0.45), and
NGC 2477 (0.0), which are shown in Figure~\ref{hrhst} superimposed on our data
(including the earlier subdwarfs).
The globular cluster data reach only the brightest of the sdM and
esdM.  The offsets however appear to be consistent with 
the esdM having abundances between -1.6 and -2.26 and the sdM 
between -1.6 and -0.6,
as deduced for our spectroscopic analysis.  SEG note that the slopes
of their CMD are steeper than the nearby field parallax subdwarfs and
attribute this to either calibration errors in the HST photometry
or real astrophysical differences.  The 
considerable difficulties in transforming HST magnitudes to 
the Cousins system are discussed by Harris { et al.} (1991) 
and Holtzman { et al.} (1995).  In particular, 
the HST F606W filter includes a substantial fraction of the $R_C$ passband.
However, the differences in slope of the main sequence 
are less evident in Figure~\ref{hrhst} 
than in the fits reported by SEG in the region of overlap.  
It appears likely that the confusion
in the definition of subdwarfs led the the inclusion of progressively
less metal-poor stars at redder colors 
in the Richer and Fahlman (1992) main sequence, in particular 
stars with V-I $\gtrsim 2.4$ which
we argue below are Intermediate Population II ($[m/H] \sim -0.6$)
and which, in any case, are clearly much fainter and redder than 
the stars used to define the globular cluster slopes in the HST study.  
The fit to the NGC 2477 lower main sequence appears to lie above the
local stars -- perhaps due to color terms but perhaps also because of the 
inclusion of binaries or non-cluster members in the fits.  

Most of the cool extreme subdwarfs in this paper come from the M92 USNO
CCD parallaxes.  Our spectroscopic metallicity scale is in agreement
with the M92 estimate that the mean $[m/H]$ of extreme subdwarfs  
is  $\sim -1.7$.   
Some authors have disputed the argument that the extreme subdwarfs represent
the 'typical' halo metallicity of $\sim -1.7$.  Richer and Fahlman (1992)
argue that the extremely hot kinematics (15 of the 17 M92 extreme subdwarfs 
have $v_{tan} > 275 {\rm ~km~s}^{-1}$ 
and the 5 stars with known total space velocities have 
$<v_{tot}> = 427 {\rm ~km~s}^{-1}$) may
point to their being more metal poor than the typical Population II stars.
However, since there is no metallicity-velocity relation within 
the higher mass stars of the Population II halo (\cite{n86}) 
this argument seems
unconvincing.  Note however that the Richer and Fahlman 
color-magnitude relation matches our sdM at V-I $\sim 1.5$
but matches the disk sequence by $V-I \sim 3$, which may be appropriate
for their sample since their ``Population II'' luminosity function 
is in fact dominated by the IPII at the faint end (\cite{reidyan}).   
Lacking a direct determination of metallicity, Eggen (1996) 
estimated values of -2.5 to -3.5  for the M92 extreme subdwarfs 
but considered these values ``guesses only.''  

\subsection{Where are the cool sdM? \label{gap}}

M92 note that their HR diagram (their Figure 10) shows a remarkable gap between
the extreme subdwarfs and the disk sequence.  In their HR diagram, only one
subdwarf appears between the two sequences (at about V-I of 2.2).  
In our sample, there is clearly a continuous distribution for 
stars with V-I$ < 2.2$, leaving the gap for $2.2 <$ V-I$ < 2.8$.
There are three major solutions to the gap, each of which was 
addressed by M92.  First, the gap may 
be an illusion caused by stellar astrophysics -- a continuous 
distribution in metallicity may not necessarily correspond to a simple
distribution in a given observational HR diagram.  Second, the gap
may be due to selection effects such that
stars in the gap are not chosen for parallax programs.  Finally,
it may represent a real lack of stars with  metallicities between the
disk sequence and the extreme subdwarf sequence. 
We will reconsider each of these options.  
It is certainly true that bandpass effects and the properties
of cool atmospheres are important in interpreting HR diagrams.  
In Figure~\ref{hrbv},
we have seen that the behavior of the B-V HR diagram is opposite
to that of the V-I diagram -- at the reddest colors, 
the subdwarfs lie above the disk sequence, and the esdM lie above the sdM.  
A second example of the
importance of bandpass effects is in the V-I HR diagram, where 
it appears that for the very coolest disk stars (LHS 2924 and 2065) 
V-I turns around, and becomes bluer with decreasing temperature  
(\cite{b91}; M92).   
Nevertheless, there is no evidence at present that the gap is 
caused by bandpass effects because
M92 report the Allard model atmospheres do not produce such an effect.
M92 consider that it is ``very unlikely'' that the gap is due to
the USNO selection criteria, and therefore offer the preliminary 
interpretation that stars with intermediate metallicities are rare
in the solar neighborhood.  Since this result is quite surprising,
we will next reconsider the selection effects that may entered this sample.

There necessarily were important selection effects in the 
catalog, since as discussed in Section~\ref{secobs}, 
even in a proper motion survey the majority of stars will not
be extremely metal poor or halo stars.  It is difficult
to know exactly what the selection effects entering the catalog were, but
early studies that emphasized the targets surely were influential.
Some of the M92 stars were never studied before (according to SIMBAD),
but others had already been noted as interesting.   
Two of their stars (LHS 3382 and LHS 489) were pointed out 
by Ake and Greenstein (1980) as being spectroscopically 
much more metal poor than other known subdwarfs.
Liebert { et al.} (1979) reported ``sdM'' spectra for five others
(LHS 192, 197, 205a, 207, and 453) from a search for nearby degenerate
stars -- their Figure 2 presents a noisy spectrum of LHS 205a with
strong hydride bands and a lack of TiO.  
Four others (as well as LHS 3382 again)
were chosen by Hartwick { et al.} (1984) for having extremely 
large reduced proper motions, defined as
$$H = M + 5 \log {{v_{tan}}/{4.74}} = m + 5 \log \mu +5 $$
The lack of the sdM is obvious in their reduced proper motion 
diagram (their Figure 5).  In contrast, there is no obvious gap in
the reduced proper motion diagram (based on B and R magnitudes
from the POSS plates) 
of the entire LHS catalog presented by Dawson (1986, his Figure 7).    
One explanation of the very large tangential velocities of the M92
sample ($<v_{tan}> = 380 {\rm ~km~s}^{-1}$) is that there 
was a implicit selection effect favoring large H.  If so, a group of
subdwarfs $\sim 1$ magnitude more luminous but the same reduced
proper motions would have $<v_{tan}> = 600 {\rm ~km~s}^{-1}$.  Since this is 
greater than the escape velocity of the galaxy ($500 {\rm ~km~s}^{-1}$, 
\cite{cll88})), these subdwarfs could not enter the sample.  
Similarly, the (unexplained) lower limit of $v_{tan} = 275 {\rm ~km~s}^{-1}$
would correspond to a lower limit of $v_{tan} = 435 {\rm ~km~s}^{-1}$.   
We therefore suggest that such a selection effect 
(as well as spectral selection of very weak TiO lines) 
would explain the lack of intermediate
metallicity stars.  As noted above, these stars are too faint 
to have been included in the more completely studied Giclas catalog
which includes many bright early sdM.
In any case, there are some likely LHS Catalog red sdM in the study of 
Hartwick { et al.} (1984) which do not yet have parallaxes  -- 
they identified both Class H stars (e.g., LHS 192, 1970, 
3259, 3382, and 3548), which we call esdM, and Class I stars 
(e.g., LHS 29 and 64) which we can identify as 
sdM stars.  They found three faint class I stars: LHS 2533, 2630, and 3189.
We confirm the latter as an sdM5.0 star.    
Further observations of these three stars would greatly increase 
the information available on cool sdM stars.

\subsection{Kinematics}

Using the trigonometric parallaxes in Table~\ref{photo}, the radial
velocities in Table~\ref{spectra}, and the LHS catalog proper motion values,
we can compute the U,V,W components of galactic velocity 
(Table~\ref{uvw}).
The errors in radial velocity, $\pm \sim 20 {\rm ~km~s}^{-1}$, 
are comparable to the uncertainties due to the trigonometric parallax errors.  
The kinematics of the sample are not representative of the Population II
due to the uncertain selection criteria, but they can provide
some information on the selection criteria used.  Mean values and standard
deviations (velocity dispersions) for the spectroscopic classes are
listed in Table~\ref{velstat}.  
The sdM, with the mean velocity component $<V> = -202 {\rm ~km~s}^{-1}$,
and the esdM, with a mean galactic velocity component
$<V> = -287 {\rm ~km~s}^{-1}$, have clear
Population II kinematics (the mean $V$ velocity for the halo 
is disputed but lies between -270 and $-180 {\rm ~km~s}^{-1}$ according to 
\cite{m93}).  Our candidate halo stars that were
classified as M V 
have both smaller $<V> = -134 {\rm ~km~s}^{-1}$
and smaller U and V velocity dispersions -- more typical
of the IPII and certainly much greater than the disk 
($<V> = -22 {\rm ~km~s}^{-1}$).  These non-subdwarfs in fact have
bandstrength indices similar to the majority of RHG nearby
stars with $V_{tan} > 100 {\rm ~km~s}^{-1}$, which should be
an IPII dominated sample.  
For  both sdM and esdM, $\sigma_U$ and $\sigma_W$ are somewhat greater
and $\sigma_V$ is somewhat less than the halo values of Norris (1986)
as expected for a sample selected by large velocity but excluding
near solar V values.   Also given in Table~\ref{velstat} are the velocity
dispersions for late M-dwarf RHG stars with ${TiO5} < 0.4$ and
either weak CaH1 ($\rm{CaH1} \ge 0.74$) or strong CaH1 ($\rm{CaH1} < 0.74$),
as well as the RHG values for nearby disk M-dwarfs.
The strong CaH1 stars have larger velocity dispersions, suggesting
that they are older and supporting the idea they are more metal poor --
but their velocities are much less than a Population II group.
Thus although CaH1 is useless as a temperature indicator for stars
cooler than M3, it may prove to be useful metallicity 
indicator for late-type (M4 to M6.5) disk stars, particularly if 
CaH features can be reproduced by future model atmospheres.  
 
We have also computed total galactocentric velocity for the stars
after correcting for the Sun's motion with respect to the 
Local Standard of Rest (\cite{mb83}) and assuming a LSR circular
velocity of $220 {\rm ~km~s}^{-1}$.  All of the stars are easily 
bound to Galaxy --
the fastest sdM (LHS 467, $v_{Gal} = 342 {\rm ~km~s}^{-1}$) and the
fastest esdM (LHS 3548, $v_{Gal} = 364 {\rm ~km~s}^{-1}$;
LHS 453, $v_{Gal} = 451 {\rm ~km~s}^{-1}$) are comfortably
less than the galactic escape speed of $\sim 500 {\rm ~km~s}^{-1}$ 
(\cite{cll88}). 

\section{Notes on Individual Stars\label{secnotes}}

\subsection{CM Dra (LHS 421)}

CM Dra (Gl 630.1) is a short period eclipsing binary with component
masses $M_A = 0.2307 \pm 0.0010$ and $M_B = 0.2136 \pm 0.0010 M_{\odot}$
and helium abundance $Y_A = 0.32$ and $Y_B = 0.31 \pm 0.04$ ( \cite{mmlt96}).
Based on the large W velocity and the observed low flaring rate
and lacking a spectroscopic metallicity,
\cite{mmlt96} interpret the system as Population II and therefore
interpret the high helium abundance as possible support for a higher
primordial helium abundance.  Using the observed spectrum (taken 
from RHG), we find that
the system not very metal poor, with a best fit model of $[m/H] = 0.0$,
and does not show any CaH excess compared to Population I stars; however,
infrared photometry (\cite{l92}) suggests it is more metal-poor than
most disk stars.  
We therefore conclude CM Dra formed from significantly enriched material.
The CNS3 W velocity of -34 
(-27 when corrected for the solar motion given in \cite{mb83}) 
implies that its maximum height above
the plane is only $\sim 350$ pc (\cite{kg89}), whereas starcounts show 
the true Population II
does not become dominant until beyond 5 kpc (\cite{m93}), so CM Dra
can probably be considered an Intermediate Population II, or
even Population I, star.  

\subsection{Barnard's Star (LHS 57) and Gl 299 (LHS 35)}

M92 have discussed the well known Barnard's Star (Gl 699) extensively.  
It lies perhaps 1.6 magnitudes below their mean disk main sequence, 
yet it is well known that spectroscopically it is an M-dwarf, a result 
that this study confirms.  We believe that the
distance below the main sequence is ``artificially'' enhanced by the
presence of the kink in the main sequence (\cite{gr96} Figure 5)
which is not fit well by a line.
At spectral type M4.5 the disk sequence, when restricted to only
single stars with high quality parallaxes and photometry,
shows a sharp 1 magnitude drop.    Hence Barnard's Star is 
only about 0.6 magnitudes below the disk sequence.  It seems likely
that Barnard's Star has a metallicity between $-0.5$ and $-1.0$.  
It is a weak outlier in TiO and CaH.  The well known star
Gl 299 (LHS 35) is quite similar in both its indices and absolute magnitude.
We identify both as Intermediate Population II stars and as cool
analogues to the spectroscopically unremarkable stars (e.g., LHS 301 and
376).   The M4.5 V star LHS 3684 which has an M92 parallax is
an only slightly cooler counterpart of Barnard's Star in
both the spectroscopic indices and the HR diagram.  

\subsection{The sdMe (LHS 482 and 2497) and Unresolved Binaries}

Two subdwarfs in this study show noticeable $H \alpha$ emission.
Gl 781 (LHS 482) shows large radial velocity
variations typical of a short period binary (\cite{j47}).
We have obtained echelle spectra that confirm the radial velocity variations
but do not show lines from the secondary (\cite{grh}).  We therefore
have not adjusted the absolute magnitude in Table\ref{photo}
for the secondary component.
Gl 455 (LHS 2497) shows weak H alpha emission (\cite{hgr96})
and our recent echelle spectra have shown the system to be a short
period double-lined spectroscopic binary (\cite{grh}).
Since the system is near-equal luminosity we have not adjusted the
color but have subtracted 0.75 from the absolute magnitude
in Table~\ref{photo}.  The system is a strong outlier
in Figure~\ref{cahtiofig}, yet even with the adjustment it
has an absolute magnitude and color similar to the subluminous
IPII dM stars.  The YPC parallax is based upon two determinations
with good agreement and has an
uncertainty of less than $10\%$.  If it is not in error, 
perhaps the chromospheric activity affects the spectrum or 
magnitude of this interesting system.  Unfortunately we have no
other examples of late sdM that is comparable to this system.
Young { et al.} (1987) argue that M-dwarf binaries with short
periods ($P \lesssim 5$ days) must show emission.  The observed radial
velocity variations are consistent with this limit.  Mass determinations 
would provide a important constraint on theories of metal poor low-mass
stars, but the short periods imply small separations ($a \lesssim 
4$ milliarcseconds) for these systems.
No other sdM or esdM show $H \alpha$ emission in this study, and
indeed no extreme M-subdwarfs with emission are yet known.

The sdMe discussed above are the only stars in the sample known
to have unresolved companions.
LHS 169, also known as Gl 129 but no longer within
the nearby star catalog (CNS3) due to an improved parallax, has been
reported to be spectroscopic binary.  Three measurements of the radial 
velocity (\cite{j47}) found radial velocities of -139.6, -91.0
and $-75.9 {\rm ~km~s}^{-1}$ with probable error $2.1 {\rm ~km~s}^{-1}$.  
The radial velocity variations have evidently never been 
confirmed or refuted since then, and Dawson and De Robertis (1988) 
suggest that the Joy (1947) velocities may be incorrect for a number of
subdwarfs.   
LHS 64 has been reported to have velocity varying between 
-292 and -242 by Joy (1947) but we have obtained 3 echelle 
spectra which are consistent with no variations at all.  Lacking any further
data, we do not apply any correction to the absolute magnitude of either star
for any companion.   Presumably at least a few of the stars in our sample
have unresolved companions that are still unknown, which could affect
the magnitudes or radial velocities in a few cases.  

\subsection{The most extreme subdwarf: LHS 453}

One important use of a classification system is identifying 
unusual stars.  Our indices have allowed us to identify 
LHS 453 as even more metal poor than the other extreme subdwarfs.
We would therefore expect this star to
be below the other stars in the HR diagram, but the M92 
observations (V-I $= 2.22$, $M_V = 13.08$) 
place the star in the midst of the  
extreme subdwarf sequence.  However, the reported mean error
in $M_V$ is still fairly large at $\pm 0.20$ magnitudes.
Perhaps coincidentally, this most metal poor star also has 
the largest galactocentric velocity in the sample, though
Dawson and De Robertis (1988) showed that despite the large tangential
velocity (M92 give $466.9 \pm 42.8$) the star is bound to the galaxy.
If the parallax is indeed reduced, the tangential velocity would also
be decreased.  Alternatively, if LHS 453 is a binary, its $M_V$ would
be fainter by up to 0.75 magnitudes.  The best fit
synthetic spectrum has $[m/H] = -2.5$, but we note that the synthetic
$[m/H] = -3.0$ spectrum is almost identical.  
Although TiO is absent, LHS 453 is not likely to be more metal poor 
than this because the observed atomic lines are prominent and 
similar in strength to those of LHS 205a
(which is only slightly later type in V-I color and CaH indices).
According to the models, the atomic lines should be much weaker or
absent for $[m/H] = -3.5$ or -4 at all temperatures.
  
\section{Conclusions\label{conclusions}}

This paper presents spectroscopy of 50 late-type stars that are significantly
more metal-poor than the Population I.  Bandstrength indices are used to
measure TiO and CaH features and show that many different metallicities
can be distinguished in the Population II M stars, as expected from
the metallicity spread seen in G-subdwarfs.  The TiO feature at $\sim 7100 \AA$
is shown to put all the extreme subdwarfs into only a few spectral subclasses
(K7 and M0).
Instead, a CaH index which reproduces the Population I spectral classification
system of Kirkpatrick, Henry, and McCarthy (1991) is used to classify
stars.  Based on the differences between the (weak) TiO and the CaH
the metal poor M stars are divided into the simple categories of M-subdwarfs
(sdM) and extreme M-subdwarfs (esdM) and spectral standards are presented
(Figures~\ref{figsdm} and~\ref{figesdm}).    

Synthetic spectra computed by Allard and Hauschildt (1995) are fit to
the observed spectra.  Metallicities of $[m/H] \sim -2$ are derived 
for the esdM and $\sim -1.2$ for the sdM.  These estimates are 
in agreement with the estimates of G89 and M92 but not those of
Richer and Fahlman (1992) or Eggen (1996).  HR diagrams based on literature
observations are presented for dM, sdM and esdM.   They are at least
qualitatively in agreement
with theoretical computations and HST observations of globular clusters.
However, the former are sensitive to the boundary conditions used and
the latter are likely to suffer from large color terms.  The agreement
of all three methods seems to indicate the metallicity scale is accurate
to $\sim 0.5$ dex.    

Given this metallicity scale, we discuss a number of interesting
subdwarfs and stars that have been called subdwarfs.  
The eclipsing binary system CM Dra is shown to be
significantly enriched such that its composition is not likely to
reflect the primordial helium abundance.  The ``old disk subdwarfs''
such as Gl 299 and Gl 699 are probably not as subluminous as
is usually thought -- taking into account the detailed structure
of the HR diagram, rather than simply fitting a straight line, they
are only $\sim 0.6$ magnitudes subluminous rather than the usually
cited $\sim 1.2$ magnitude.  This naturally explains the only slight spectral
differences, which are shown to be similar to differences of 
hotter M-dwarfs that are also $\sim 0.6$ magnitudes subluminous. 
Finally, we show that 
LHS 453 is more metal poor than the other extreme subdwarfs with 
$[m/H] \sim -2.5$ to -3.  

\acknowledgments

I would like to thank Neill Reid for helpful comments 
and help with the observations.    
The referee, Todd Henry, provided comments which substantially improved
this paper. 
I am indebted to France Allard and Peter Hauschildt
for making their synthetic spectra available.
I am grateful for Greenstein and Kinsley Fellowships and
partial support through NASA grant GO-5353.0-93A. 
This research has made use of the Simbad database, operated at
CDS, Strasbourg, France.

\begin{deluxetable}{lccc}
\tablecaption{Spectroscopic Indices}
\label{index} 
\tablewidth{0pt}
\tablenum{1}
\tablehead{
\colhead{Band} & \colhead{S1} & \colhead{W} & \colhead{S2}
}
\startdata
TiO 5  & 7042-7046 &7126-7135 & \nl
CaH 1  & 6345-6355 &6380-6390 &6410-6420 \nl
CaH 2  & 7042-7046 &6814-6846\nl
CaH 3  & 7042-7046 &6960-6990 \nl
\enddata
\end{deluxetable}


\begin{deluxetable}{rlrrrrrcr}
\tablewidth{0pc}
\tablenum{2}
\tablecaption{Spectroscopic Observations}
\label{spectra}
\tablehead{
\colhead{LHS } & 
\colhead{Name} &
\colhead{TiO5} & 
\colhead{CaH1} & 
\colhead{CaH2} & 
\colhead{CaH3} & 
\colhead{$v_{rad}$} &
\colhead{class} &
\colhead{Source} }
\startdata

 12 &  G 003-036  & 0.894 & 0.852 & 0.746 & 0.878 & 25.1 & sdM0.0 & 60 \nl 
 20 &  GJ 1062   & 0.677 & 0.708 & 0.499 & 0.710 & -104.3 & sdM2.5 & RHG \nl 
 29 &  Kapteyn's  & 0.810 & 0.778 & 0.594 & 0.792 & 242.8 & sdM1.0 & 100 \nl 
 42 &  Ross 451  & 0.925 & 0.871 & 0.757 & 0.885 & -148.0 & sdM0.0 & 60 \nl 
 55 &  GJ 1200  & 0.439 & 0.816 & 0.420 & 0.705 & -21.2 & M3.5 V & 60 \nl 
 57 &  Barnard's  & 0.394 & 0.752 & 0.394 & 0.652 & -120.2 & M4.0 V & 60 \nl 
 61 &  Gl 817  & 0.645 & 0.823 & 0.582 & 0.799 & 25.9 & M1.5 V & 60 \nl 
 64 &  Wo 9722   & 0.773 & 0.731 & 0.572 & 0.754 & -235.5 & sdM1.5 & RHG \nl 
 104 &  G 030-048  & 1.009 & 0.911 & 0.854 & 0.938 & -173.8 & esdK7 & 60 \nl 
 156 &  G 004-029  & 0.655 & 0.648 & 0.445 & 0.659 & -42.5 & sdM3.0 & 100 \nl 
 161 &  G 075-047  & 0.964 & 0.739 & 0.515 & 0.751 & -35.1 & esdM2.0 & 100 \nl 
 169 &  G 005-022  & 1.008 & 0.879 & 0.862 & 0.951 & -119.2 & esdK7 & 60 \nl 
 170 &  G 078-026  & 1.007 & 0.986 & 0.931 & 0.961 & -183.2 & sdK & 60 \nl 
 173 &  G 038-001  & 0.961 & 0.942 & 0.868 & 0.938 & -159.2 & sdK7  & 60 \nl 
 174 &  G 037-040  & 0.877 & 0.817 & 0.686 & 0.833 & -225.2 & sdM0.5 & 60 \nl 
 178 &  G 079-059  & 0.674 & 0.766 & 0.551 & 0.760 & -59.7 & sdM1.5 & 60 \nl 
 182 &  G 095-059  & 0.975 & 0.842 & 0.735 & 0.843 & -238.1 & esdM0.0 & 60 \nl 
 185 &  G 007-017  & 0.957 & 0.737 & 0.659 & 0.795 & 37.7 & esdM0.5 & 60 \nl 
 192 &  LP 302-31  & 0.946 & 0.710 & 0.603 & 0.768 & 79.9 & esdM1.0 & 200 \nl 
 205a &  LP 417-44  & 0.889 & 0.512 & 0.347 & 0.513 & -76.4 & esdM4.5 & 200 \nl 
 211 &  G 099-033  & 0.796 & 0.797 & 0.703 & 0.837 & -129.2 & sdM0.0 & 100 \nl 
 216 &  G 105-023  & 0.725 & 0.743 & 0.496 & 0.732 & 392.5 & sdM2.0 & 60 \nl 
 218 &  G 103-046  & 0.500 & 0.759 & 0.413 & 0.686 & -20.4 & M3.5 V & 100 \nl 
 236 &  G 251-044  & 0.980 & 0.944 & 0.879 & 0.931 & 66.8 & sdK7  & 60 \nl 
 254 &  LP 666-11  & 0.178 & 0.678 & 0.230 & 0.517 & 20.2 & M6.5 V & 100 \nl 
 272 &  LP 788-27  & 0.678 & 0.676 & 0.441 & 0.669 & 264.4 & sdM3.0 & 100 \nl 
 276 &  G 117-061  & 0.686 & 0.888 & 0.623 & 0.828 & 46.9 & M1.0 V & 60 \nl 
 301 &  GJ 1146   & 0.481 & 0.773 & 0.431 & 0.683 & 94.4 & M3.5 V & 60 \nl 
 307 &  G 176-040  & 0.910 & 0.763 & 0.671 & 0.837 & -64.8 & sdM0.5 & 60 \nl 
 320 &  G 011-035  & 0.620 & 0.756 & 0.512 & 0.751 & 81.1 & sdM2.0 & 60 \nl 
 343 &  G 061-021  & 1.001 & 0.953 & 0.901 & 0.937 & 165.5 & sdK & 60 \nl 
 364 &  G 165-047  & 0.979 & 0.699 & 0.604 & 0.740 & 12.8 & esdM1.5 & 60 \nl 
 375 &  LP 857-48  & 0.829 & 0.499 & 0.372 & 0.547 & 181.2 & esdM4.0 & 60 \nl 
 376 &  G 135-067  & 0.448 & 0.795 & 0.423 & 0.682 & -19.6 & M3.5 V & 60 \nl 
 377 &  LP 440-52  & 0.232 & 0.567 & 0.205 & 0.396 & 179.7 & sdM7.0 & 200 \nl 
 407 &  LP 803-27  & 0.601 & 0.584 & 0.351 & 0.528 & -173.1 & sdM5.0 & 60 \nl 
 410 &  G 016-018  & 0.724 & 0.822 & 0.612 & 0.814 & 10.5 & M1.0 V & 60 \nl 
 418 &  G 138-025  & 0.941 & 0.908 & 0.814 & 0.910 & -14.4 & K7 V & 60 \nl 
 425 &  G 138-059  & 0.646 & 0.762 & 0.503 & 0.738 & -49.3 & sdM2.0 & 60 \nl 
 453 &  LP 139-14  & 1.070 & 0.514 & 0.407 & 0.590 & 32.1 & esdM3.5 & 200 \nl 
 460 &  GJ 1225   & 0.337 & 0.736 & 0.321 & 0.603 & -53.2 & M5.0 V & 60 \nl 
 467 &  G 021-023  & 0.966 & 0.898 & 0.786 & 0.897 & 183.6 & sdK7 & 60 \nl 
 479 &  G 142-052  & 0.792 & 0.768 & 0.610 & 0.801 & -75.4 & sdM1.0 & 60 \nl 
 482 &  Gl 781  & 0.774 & 0.718 & 0.558 & 0.746 & -107.6 & sdM1.5 & RHG \nl 
 489 &  LP515-3  & 0.998 & 0.853 & 0.766 & 0.889 & -95.3 & esdM0.0 & 60 \nl 
 491 &  G 210-019  & 0.843 & 0.708 & 0.582 & 0.759 & -142.9 & sdM1.5 & 60 \nl 
 522 &  G 018-051  & 1.024 & 0.926 & 0.854 & 0.929 & -159.8 & esdK7 & 60 \nl 
 536 &  G 128-034  & 0.883 & 0.784 & 0.659 & 0.826 & -57.8 & sdM0.5 & 60 \nl 
 537 &  G 028-043  & 1.002 & 1.010 & 0.978 & 0.982 & -113.0 & K5 V & 60 \nl 
 1088 &  G 217-055  & 0.454 & 0.781 & 0.427 & 0.740 & 17.5 & M3.5 V & 60 \nl 
 1164 &  L 220-27  & 0.654 & 0.868 & 0.988 & 0.955 & -22.5 & K5 V & 100 \nl 
 1174 &  LP 406-47  & 0.924 & 0.607 & 0.454 & 0.659 & -111.7 & esdM3.0 & 200\tablenotemark{a} \nl 
 1481 &  LP 711-32  & 0.554 & 0.708 & 0.443 & 0.681 & 94.5 & sdM3.0 & RHG \nl 
 1742a &  LP 417-42  & 0.689 & 0.428 & 0.284 & 0.429 & 218.1 & esdM5.5 & 200 \nl 
 1970 &  LP 484-6  & 0.878 & 0.589 & 0.475 & 0.657 & 9.8 & esdM2.5 & 200 \nl 
 2110 &  LP 787-4  & 0.362 & 0.747 & 0.366 & 0.653 & 37.8 & M4.0 V & 100 \nl 
 2497 &  Gl 455  & 0.542 & 0.656 & 0.400 & 0.627 & 36.2 & sdM3.5 & RHG \nl 
 2715 &  Gl 506.1  & 0.994 & 1.019 & 0.971 & 0.975 & 36.0 & sdK & 60 \nl 
 2852 &  LP 856-36  & 0.651 & 0.736 & 0.507 & 0.725 & 13.3 & sdM2.0 & RHG \nl 
 3061 &  LP 502-32  & 0.737 & 0.497 & 0.292 & 0.497 & 65.4 & esdM5.0 & 200 \nl 
 3073 &  G 137-008  & 0.989 & 0.918 & 0.869 & 0.922 & -254.5 & sdK7 & 60 \nl 
 3084 &  G 015-026  & 0.840 & 0.878 & 0.724 & 0.857 & -43.8 & sdK & 60 \nl 
 3189 &  LP 225-22  & 0.341 & 0.637 & 0.305 & 0.579 & -106.2 & sdM5.0 & 200 \nl 
 3192 &  Gl 871  & 0.454 & 0.768 & 0.433 & 0.706 & 0.6 & M3.5 V & 60 \nl 
 3193 &  G 169-007  & 0.688 & 0.878 & 0.635 & 0.829 & 20.1 & M1.0 V & 60 \nl 
 3259 &  LP 686-36  & 0.995 & 0.811 & 0.657 & 0.803 & -217.0 & esdM0.5 & 200 \nl 
 3382 &  LP 24-219  & 0.937 & 0.635 & 0.485 & 0.667 & -138.9 & esdM2.5 & 200\tablenotemark{a} \nl 
 3409 &  LP 141-1  & 0.443 & 0.645 & 0.355 & 0.587 & -80.3 & sdM4.5 & RHG \nl 
 3480 &  LP 869-24  & 0.458 & 0.695 & 0.363 & 0.632 & 12.9 & sdM4.0 & 200 \nl 
 3481 &  LP 753-21  & 0.782 & 0.669 & 0.467 & 0.694 & 75.4 & sdM2.5 & 200 \nl 
 3548 &  LP 695-96  & 0.886 & 0.839 & 0.435 & 0.660 & 80.6 & esdM3.0 & 200\tablenotemark{a} \nl 
 3628 &  LP 757-13  & 0.908 & 0.643 & 0.576 & 0.742 & 10.1 & esdM1.5 & 100 \nl 
 3684 &  LP 518-12  & 0.377 & 0.733 & 0.344 & 0.637 & -75.3 & M4.5 V & 200 \nl 
 3867 &  G 067-030  & 1.024 & 0.997 & 0.962 & 0.982 & -108.8 & sdK & 60 \nl 
 3957 &  G 190-026  & 0.727 & 0.846 & 0.648 & 0.842 & -85.0 & M0.5 V & 60 \nl 
 6304 &  G 169-009  & 0.587 & 0.785 & 0.505 & 0.751 & -9.1 & M2.5 V & 60 \nl 
 \nodata & G 017-006 & 0.781 & 0.920 & 0.702 & 0.875 & 8.9 & M0.0 V & 60 \nl 
 \nodata & Sm 183 & 0.880 & 0.947 & 0.867 & 0.920 & -12.3 & gM & 100 \nl 
 \nodata & TVLM 832-42500 & 0.186 & 0.842 & 0.171 & 0.452 & 72.1 & $>$M6.0 V & 200 \nl 

\tablenotetext{a}{Observed with August 1995 setup}

\enddata
\end{deluxetable}

\begin{deluxetable}{lccclccclcccr}
\scriptsize
\tablewidth{0pt}
\tablenum{3}
\tablecaption{Model Best Fits}
\label{fit}
\tablehead{
\colhead{Star } & 
\colhead{Class} & 
\colhead{$T_{eff}$} & 
\colhead{$[m/H]$} & 
\colhead{Star} & 
\colhead{Class} & 
\colhead{$T_{eff}$} & 
\colhead{$[m/H]$} & 
\colhead{Star} & 
\colhead{Class} & 
\colhead{$T_{eff}$} & 
\colhead{$[m/H]$} & 
}
\startdata

  LHS  410  &   M1.0 V   &  3700  & -1.0  &  LHS  307  &   sdM0.5   &  3700  & -1.5  &  LHS  3259  &   esdM0.5   &  3700  & -2.0  &  \nl 
  LHS  301  &   M3.5 V   &  3700  & -0.5  &  LHS  29  &   sdM1.0   &  3700  & -1.5  &  LHS  192  &   esdM1.0   &  3600  & -2.5  &  \nl 
  LHS  218  &   M3.5 V   &  3600  & -1.0  &  LHS  479  &   sdM1.0   &  3700  & -1.0  &  LHS  364  &   esdM1.5   &  3600  & -2.5  &  \nl 
  LHS  3192  &   M3.5 V   &  3600  & -0.5  &  LHS  178  &   sdM1.5   &  3600  & -1.0  &  LHS  3628  &   esdM1.5   &  3600  & -1.5  &  \nl 
  LHS  376  &   M3.5 V   &  3600  & -0.5  &  LHS  482  &   sdM1.5   &  3600  & -1.0  &  LHS  161  &   esdM2.0   &  3600  & -2.0  &  \nl 
  LHS  6304  &   M2.5 V   &  3600  & -1.0  &  LHS  491  &   sdM1.5   &  3600  & -1.5  &  LHS  3382  &   esdM2.5   &  3500  & -1.5  &  \nl 
  LHS  1088  &   M3.5 V   &  3500  &  0.0  &  LHS  64  &   sdM1.5   &  3600  & -1.0  &  LHS  1970  &   esdM2.5   &  3500  & -1.5  &  \nl 
  LHS  2110  &   M4.0 V   &  3500  &  0.0  &  LHS  320  &   sdM2.0   &  3600  & -1.0  &  LHS  1174  &   esdM3.0   &  3500  & -1.5  &  \nl 
  LHS  55  &   M3.5 V   &  3500  &  0.0  &  LHS  216  &   sdM2.0   &  3600  & -1.0  &  LHS  3548  &   esdM3.0   &  3500  & -1.5  &  \nl 
  LHS  57  &   M4.0 V   &  3500  & -0.5  &  LHS  425  &   sdM2.0   &  3600  & -1.0  &  LHS  453  &   esdM3.5   &  3400  & -2.5  &  \nl 
  LHS  3684  &   M4.5 V   &  3500  & -0.5  &  LHS  2852  &   sdM2.0   &  3600  & -1.0  &  LHS  375  &   esdM4.0   &  3400  & -2.0  &  \nl 
  LHS  460  &   M5.0 V   &  3300  &  0.0  &  LHS  20  &   sdM2.5   &  3600  & -1.0  &  LHS  205a  &   esdM4.5   &  3400  & -2.0  &  \nl 
  LHS  254  &   M6.5 V   &  3200  &  0.0  &  LHS  3481  &   sdM2.5   &  3600  & -1.5  &  LHS  3061  &   esdM5.0   &  3300  & -2.0  &  \nl 
  TVLM 832-42500  &   $>$M6.0 V   &  3000  &  0.0  &  LHS  272  &   sdM3.0   &  3600  & -1.5  &  LHS  1742a  &   esdM5.5   &  3300  & -2.0  &  \nl 
   &  &  &  &  LHS  156  &   sdM3.0   &  3600  & -1.5  &   &  &  &  &  \nl 
   &  &  &  &  LHS  1481  &   sdM3.0   &  3600  & -1.0  &   &  &  &  &  \nl 
   &  &  &  &  LHS  2497  &   sdM3.5   &  3500  & -1.0  &   &  &  &  &  \nl 
   &  &  &  &  LHS  3480  &   sdM4.0   &  3500  & -1.0  &   &  &  &  &  \nl 
   &  &  &  &  LHS  3409  &   sdM4.5   &  3400  & -1.0  &   &  &  &  &  \nl 
   &  &  &  &  LHS  3189  &   sdM5.0   &  3400  & -1.0  &   &  &  &  &  \nl 
   &  &  &  &  LHS  407  &   sdM5.0   &  3400  & -1.5  &   &  &  &  &  \nl 
   &  &  &  &  LHS  377  &   sdM7.0   &  3200  & -1.5  &   &  &  &  &  \nl 
\enddata
\end{deluxetable}

\begin{deluxetable}{rccccccccrl}
\tablewidth{0pt}
\tablenum{4}
\tablecaption{Photometry}
\label{photo}
\tablehead{
\colhead{LHS} & 
\colhead{V} & 
\colhead{B-V} & 
\colhead{V-R} & 
\colhead{R-I} & 
\colhead{V-I} & 
\colhead{$\pi$ (\arcsec)} & 
\colhead{$\sigma_{\pi}$ (\arcsec)} & 
\colhead{$\Delta_{LK}$\tablenotemark{a}} &
\colhead{$M_V$} &
\colhead{Source\tablenotemark{b}} }
\startdata
 12 & 12.24 & 1.45 & 0.90 & 0.90 & 1.80 & 0.0359 & 0.0031 &  -0.08 & 10.02 & Y,L \nl 
 20 & 13.01 & 1.68 & 1.03 & 1.17 & 2.20 & 0.0648 & 0.0025 &  -0.01 & 12.07 & Y,L \nl 
 29 & 8.84 & 1.56 & 0.96 & 1.00 & 1.96 & 0.2583 & 0.0065 &  -0.01 & 10.90 & Y,L \nl 
 35 & 12.83 & 1.72 & 1.25 & 1.67 & 2.92 & 0.1480 & 0.0022 &  -0.00 & 13.68 & Y,L \nl 
 42 & 12.23 & 1.45 & 1.06 & 0.92 & 1.98 & 0.0327 & 0.0025 &  -0.06 & 9.80 & Y,E79 \nl 
 44 & 6.45 & 0.76 & \nodata & \nodata & \nodata & 0.1127 & 0.0014 &  -0.00 & 6.71 & Y,S \nl 
 52 & 9.43 & 0.84 & 0.52 & 0.49 & 1.01 & \nodata & \nodata & \nodata & \nodata & R
 \nl 
 53 & 9.04 & 0.78 & 0.45 & 0.46 & 0.91 & \nodata & \nodata & \nodata & \nodata & R
 \nl 
 55 & 12.90 & 1.54 & \nodata & \nodata & \nodata & 0.0550 & 0.0029 &  -0.03 & 11.60 & HD,S \nl 
 64 & 13.30 & 1.55 & \nodata & \nodata & \nodata & 0.0419 & 0.0022 &  -0.03 & 11.41 & Y,S \nl 
 103 & 14.16 & \nodata & 1.15 & 1.47 & 2.62 & 0.0536 & 0.0021 &  -0.02 & 12.81 & Y,WU \nl 
 104 & 13.78 & 1.34 & 0.81 & 0.91 & 1.72 & \nodata & \nodata & \nodata & \nodata & E79
 \nl 
 137 & 13.36 & 1.60 & \nodata & \nodata & \nodata & 0.0463 & 0.0027 &  -0.03 & 11.69 & Y,S \nl 
 139 & 15.05 & 1.86 & 1.24 & 1.61 & 2.85 & 0.0485 & 0.0044 &  -0.09 & 13.48 & Y,E79 \nl 
 156 & 14.89 & 1.70 & 0.92 & 1.25 & 2.17 & 0.0277 & 0.0035 &  -0.18 & 12.1 & Y,E79 \nl 
 161 & 14.75 & 1.55 & 1.01 & 0.96 & 1.98 & 0.0260 & 0.0048 &  -0.42 & 11.82 & Y,E79 \nl 
 163 & 13.06 & 1.57 & 1.12 & 1.43 & 2.55 & 0.0496 & 0.0043 &  -0.08 & 11.54 & Y,B \nl 
 169 & 14.13 & 1.45 & 0.91 & 0.76 & 1.72 & 0.0309 & 0.0023 &  -0.06 & 11.58 & Y,L,E79 \nl 
 170 & 10.69 & 1.22 & 0.80 & 0.70 & 1.50 & 0.0348 & 0.0025 &  -0.05 & 8.40 & Y,E79 \nl 
 173 & 11.12 & 1.31 & \nodata & \nodata & \nodata & 0.0353 & 0.0031 &  -0.08 & 8.86 & Y,S \nl 
 174 & 12.75 & 1.52 & 0.94 & 0.89 & 1.83 & 0.0204 & 0.0045 &  -0.65 & 9.30 & Y,DF89 \nl 
 175 & 9.91 & 0.65 & \nodata & \nodata & \nodata & 0.0218 & 0.0022 &  -0.11 & 6.6 & Y,R \nl 
 178 & 12.90 & 1.55 & 0.95 & 1.11 & 2.06 & 0.0451 & 0.0066 &  -0.24 & 11.17 & Y,E79 \nl 
 182 & 13.42 & 1.57 & \nodata & \nodata & \nodata & 0.0234 & 0.0025 &  -0.12 & 10.27 & Y,S \nl 
 185 & 15.30 & 1.79 & 0.98 & 0.85 & 1.83 & 0.0167 & 0.0046 & L & 11.41 & Y,DF89 \nl 
 192 & 17.33 & \nodata & \nodata & \nodata & 1.98 & 0.0102 & 0.0008 &  -0.06 & 12.37 & M92 \nl 
 205a & 18.93 & \nodata & \nodata & \nodata & 2.35 & 0.0104 & 0.0013 &  -0.17 & 14.02 & M92 \nl 
 211 & 14.11 & 1.45 & 0.91 & 1.05 & 1.96 & 0.0188 & 0.0029 &  -0.28 & 10.48 & Y,L \nl 
 216 & 14.66 & 1.62 & 0.99 & 1.09 & 2.08 & 0.0306 & 0.0030 &  -0.10 & 12.09 & Y,L \nl 
 218 & 14.84 & 1.57 & 1.12 & 1.33 & 2.45 & 0.0296 & 0.0030 &  -0.11 & 12.20 & Y,DF89 \nl 
 232 & 13.72 & 1.15 & \nodata & \nodata & \nodata & 0.0148 & 0.0025 &  -0.34 & 9.57 & Y,S \nl 
 236 & 13.10 & 1.33 & \nodata & \nodata & \nodata & 0.0187 & 0.0018 &  -0.10 & 9.46 & Y,S \nl 
 241 & 8.32 & 0.62 & \nodata & \nodata & \nodata & \nodata & \nodata & \nodata & \nodata & S
 \nl 
 254 & 17.41 & 1.75 & 1.90 & 2.07 & 3.97 & \nodata & \nodata & \nodata & \nodata & L
 \nl 
 272 & 13.46 & \nodata & 1.34 & 1.36 & 2.70 & \nodata & \nodata & \nodata & \nodata & DF89
 \nl 
 276 & 11.86 & 1.43 & 0.93 & 1.04 & 1.97 & 0.0347 & 0.0025 &  -0.05 & 9.56 & Y,B \nl 
 301 & 13.57 & 1.57 & 1.08 & 1.39 & 2.47 & 0.0540 & 0.0036 &  -0.04 & 12.23 & Y,W88,E79 \nl 
 307 & 15.22 & 1.56 & \nodata & \nodata & 1.92 & 0.0183 & 0.0027 &  -0.25 & 11.53 & Y,L* \nl 
 320 & 14.00 & 1.54 & 0.88 & 1.26 & 2.14 & 0.0260 & 0.0034 &  -0.19 & 11.07 & Y,E79 \nl 
 343 & 13.87 & 1.30 & 0.87 & 0.69 & 1.56 & 0.0167 & 0.0031 &  -0.43 & 9.98 & HD,L \nl 
 364 & 14.61 & 1.71 & 1.03 & 0.92 & 1.95 & 0.0374 & 0.0037 &  -0.10 & 12.47 & Y,DF92 \nl 
 375 & 15.68 & 1.87 & 1.08 & 1.12 & 2.20 & 0.0395 & 0.0010 &  -0.01 & 13.66 & RA \nl 
 376 & 15.00 & 1.64 & 1.08 & 1.33 & 2.41 & 0.0207 & 0.0048 &  -0.74 & 11.58 & Y,E79 \nl 
 377 & 18.39 & \nodata & \nodata & \nodata & 3.48 & 0.0284 & 0.0008 &  -0.01 & 15.66 & M92 \nl 
 407 & 16.57 & 1.93 & 1.06 & 1.33 & 2.39 & 0.0315 & 0.0020 &  -0.04 & 14.06 & RA \nl 
 410 & 13.36 & \nodata & 0.94 & 1.06 & 2.00 & 0.0240 & 0.0034 &  -0.23 & 10.26 & Y,HD,W91 \nl 
 418 & 13.51 & 1.42 & 0.88 & 0.80 & 1.68 & 0.0141 & 0.0033 &  -0.76 & 9.26 & HD,La \nl 
 420 & 7.30 & 0.54 & \nodata & \nodata & \nodata & 0.0324 & 0.0057 &  -0.37 & 4.85 & Y,S \nl 
 425 & 15.00 & 1.60 & \nodata & \nodata & \nodata & 0.0261 & 0.0047 &  -0.40 & 12.08 & Y,S \nl 
 453 & 18.02 & \nodata & \nodata & \nodata & 2.22 & 0.0103 & 0.0009 &  -0.08 & 13.08 & M92 \nl 
 460 & 15.40 & 1.77 & 1.34 & 1.70 & 3.04 & 0.0544 & 0.0028 &  -0.03 & 14.08 & Y,DF92 \nl 
 467 & 12.21 & 1.43 & \nodata & \nodata & 1.70 & 0.0362 & 0.0020 &  -0.03 & 10.00 & Y,L \nl 
 479 & 14.31 & 1.51 & \nodata & \nodata & \nodata & 0.0224 & 0.0023 &  -0.11 & 11.06 & Y,S \nl 
 482 & 11.98 & 1.56 & 0.95 & 1.04 & 1.99 & 0.0603 & 0.0017 &  -0.01 & 10.88 & Y,L \nl 
 489 & 15.48 & 1.69 & 0.91 & 0.86 & 1.77 & 0.0189 & 0.0036 &  -0.45 & 11.86 & Y,DF89    \nl 
 491 & 14.70 & 1.68 & 0.98 & 0.98 & 1.96 & 0.0211 & 0.0036 &  -0.35 & 11.32 & Y,DF92 \nl 
 522 & 14.15 & 1.41 & 0.84 & 0.78 & 1.62 & 0.0268 & 0.0021 &  -0.06 & 11.29 & Y,B \nl 
 536 & 14.65 & 1.55 & \nodata & \nodata & \nodata & 0.0227 & 0.0025 &  -0.13 & 11.43 & Y,S \nl 
 537 & 9.96 & 0.70 & \nodata & \nodata & \nodata & \nodata & \nodata & \nodata & \nodata & S
 \nl 
 1174 & 16.99 & \nodata & \nodata & \nodata & 2.09 & 0.0157 & 0.0012 &  -0.06 & 12.97 & M92 \nl 
 1319 & 14.82 & 1.45 & \nodata & \nodata & \nodata & \nodata & \nodata & \nodata & \nodata & S
 \nl 
 1481 & 12.67 & 1.73 & 1.08 & 1.32 & 2.40 & \nodata & \nodata & \nodata & \nodata & CNS3,E87
 \nl 
 1555 & 13.90 & 1.70 & \nodata & \nodata & \nodata & 0.0168 & 0.0019 &  -0.14 & 10.03 & Y,S \nl 
 1742a & 18.80 & \nodata & \nodata & \nodata & 2.74 & 0.0134 & 0.0012 &  -0.08 & 14.44 & M92 \nl 
 1970 & 17.76 & 1.68 & \nodata & \nodata & 2.09 & 0.0129 & 0.0008 &  -0.04 & 13.31 & M,L \nl 
 2045 & 18.49 & \nodata & \nodata & \nodata & 2.46 & 0.0111 & 0.0009 &  -0.07 & 13.72 & M92 \nl 
 2110 & 16.97 & \nodata & \nodata & \nodata & 2.84 & 0.0151 & 0.0010 &  -0.04 & 12.86 & M92 \nl 
 2204 & 16.80 & \nodata & \nodata & \nodata & 3.23 & 0.0288 & 0.0009 &  -0.01 & 14.10 & M92 \nl 
 2497 & 12.85 & 1.74 & 1.08 & 1.22 & 2.47 & 0.0493 & 0.0031 &  -0.04 & 12.06\tablenotemark{c} & Y,S \nl 
 2715 & 10.83 & 1.03 & 0.62 & 0.55 & 1.17 & 0.0286 & 0.0033 &  -0.14 & 8.11 & Y,A,AM,B \nl 
 2852 & 12.15 & 1.70 & 1.06 & 1.22 & 2.18 & 0.0395 & 0.0174 & L & 10.13 & Y,CNS3,E87 \nl 
 3061 & 19.50 & \nodata & \nodata & \nodata & 2.61 & 0.0089 & 0.0008 &  -0.08 & 14.25 & M92 \nl 
 3073 & 13.69 & 1.41 & 0.84 & 0.74 & 1.58 & 0.0187 & 0.0034 &  -0.41 & 10.05 & Y,DF92 \nl 
 3084 & 13.43 & 1.43 & 0.93 & 0.90 & 1.83 & 0.0190 & 0.0025 &  -0.19 & 9.82 & Y,DF92 \nl 
 3181 & 17.18 & 1.69 & 1.02 & 1.19 & 2.21 & 0.0265 & 0.0030 &  -0.14 & 14.3 & RA \nl 
 3189 & 18.10 & \nodata & 1.11 & 1.70 & 2.81 & \nodata & \nodata & \nodata & \nodata & HCM
 \nl 
 3192 & 14.76 & 1.63 & 1.11 & 1.38 & 2.49 & 0.0274 & 0.0023 &  -0.07 & 11.95 & Y, \nl 
 3193 & 12.49 & 1.43 & 0.93 & 0.99 & 1.92 & 0.0274 & 0.0023 &  -0.07 & 9.68 & Y,DF92 \nl 
 3259 & 18.26 & \nodata & \nodata & \nodata & 2.03 & 0.0064 & 0.0012 &  -0.44 & 12.29 & M92 \nl 
 3382 & 17.02 & 1.99 & 1.03 & 1.06 & 2.09 & 0.0104 & 0.0009 &  -0.08 & 12.11 & M,L \nl 
 3409 & 15.16 & 1.85 & 1.21 & 1.61 & 2.82 & 0.0500 & 0.0013 &  -0.01 & 13.65 & Y,M92 \nl 
 3480 & 17.24 & \nodata & \nodata & \nodata & 2.76 & 0.0177 & 0.0008 &  -0.02 & 13.48 & M92 \nl 
 3481 & 17.58 & \nodata & \nodata & \nodata & 2.35 & 0.0067 & 0.0007 &  -0.12 & 11.71 & M92 \nl 
 3548 & 17.52 & 1.78 & \nodata & \nodata & 2.09 & 0.0083 & 0.0006 &  -0.05 & 12.12 & M,L \nl 
 3555 & 17.93 & 1.94 & 1.10 & 0.97 & 2.07 & 0.0125 & 0.0060 & L & 13.41 & RA \nl 
 3628 & 17.41 & \nodata & \nodata & \nodata & 2.05 & 0.0088 & 0.0008 &  -0.09 & 12.13 & M92 \nl 
 3684 & 17.95 & \nodata & \nodata & \nodata & 2.92 & 0.0152 & 0.0010 &  -0.04 & 13.86 & M92 \nl 
 3867 & 13.41 & 1.19 & \nodata & \nodata & \nodata & 0.0120 & 0.0050 & L & 8.81 & Y,S \nl 
 3957 & 13.55 & 1.39 & \nodata & \nodata & \nodata & 0.0257 & 0.0037 &  -0.23 & 10.6 & Y,S \nl 
 4037 & 13.52 & 0.88 & \nodata & 0.47 & \nodata & \nodata & \nodata & \nodata & \nodata & R89
 \nl 
 6304 & 14.07 & \nodata & 1.05 & 1.25 & 2.30 & \nodata & \nodata & \nodata & \nodata & W84 
 \nl 
\enddata
\tablenotetext{a} {L indicates the Lutz-Kelker correction is undefined 
but $< -0.80$.}
\tablenotetext{b} {References for the trigonometric parallaxes and 
photometry are:  
B - Bessell (1990),
CNS3 - Gliese and Jahreiss (1991),
DF89 - Dawson and Forbes (1989), 
DF92 - Dawson and Forbes (1992),
E79 - Eggen (1979), 
E87 - Eggen(1987), 
HD - Harrington and Dahn (1980),
L - Leggett (1992),
M92 - Monet {\it et al.} (1992), 
R - Ryan (1992),
RA - Ruiz and Anguita (1993),
S - SIMBAD,
T - Tinney {\it et al.} (1995),
W84 - Weis (1984),
W88 - Weis (1987),
W91 - Weis (1991),
WU - Weis and Upgren (1982),
Y - Van Altena {\it et al.} (1991)
}
\tablenotetext{c}{Includes 0.75 magnitude correction for unresolved companion}
\end{deluxetable}

\begin{deluxetable}{lrrrrlrrrr}
\tablewidth{0pt}
\tablenum{5}
\tablecaption{Space Velocities}
\label{uvw}
\tablehead{
\colhead{LHS } & 
\colhead{U} & 
\colhead{V} & 
\colhead{W} &
\colhead{$v_{tan}$} &
\colhead{LHS } & 
\colhead{U} & 
\colhead{V} & 
\colhead{W} &
\colhead{$v_{tan}$}  
}
\startdata
  LHS   12 &  -196 &  -254 &    15 &   321  &  LHS  453 &  -239 &  -148 &   371 &   465  \nl 
  LHS   20 &   140 &  -185 &    76 &   221  &  LHS  460 &   151 &   -48 &    20 &   151  \nl 
  LHS   29 &    19 &  -284 &   -53 &   159  &  LHS  467 &   282 &   -90 &  -117 &   260  \nl 
  LHS   42 &   178 &  -416 &   180 &   465  &  LHS  479 &   189 &  -260 &   -17 &   313  \nl 
  LHS   55 &   -92 &  -100 &   112 &   175  &  LHS  482 &   103 &  -116 &    19 &   114  \nl 
  LHS   57 &  -148 &     0 &    16 &    89  &  LHS  489 &   147 &  -257 &  -116 &   304  \nl 
  LHS   61 &   121 &  -141 &    -3 &   184  &  LHS  491 &   245 &  -210 &    70 &   298  \nl 
  LHS   64 &   259 &  -190 &   -88 &   237  &  LHS  522 &    30 &  -318 &   -83 &   289  \nl 
  LHS  156 &     1 &  -204 &   -42 &   204  &  LHS  522 &    30 &  -318 &   -83 &   289  \nl 
  LHS  161 &   -84 &  -236 &    94 &   265  &  LHS  536 &  -268 &  -121 &   -31 &   290  \nl 
  LHS  169 &    32 &  -273 &    51 &   253  &  LHS 1088 &  -175 &   -57 &   114 &   216  \nl 
  LHS  170 &    96 &  -234 &   -13 &   175  &  LHS 1174 &   -88 &  -286 &   -19 &   278  \nl 
  LHS  173 &    56 &  -255 &    16 &   208  &  LHS 1742a &  -229 &  -263 &   108 &   293  \nl 
  LHS  174 &   139 &  -382 &  -127 &   362  &  LHS 1970 &   158 &  -318 &   -49 &   359  \nl 
  LHS  178 &    60 &  -165 &    -6 &   165  &  LHS 2110 &   154 &  -103 &     4 &   181  \nl 
  LHS  182 &   123 &  -335 &  -118 &   291  &  LHS 2497 &   -71 &   -38 &    21 &    76  \nl 
  LHS  185 &     4 &  -289 &  -169 &   333  &  LHS 2715 &   -57 &   -94 &    28 &   108  \nl 
  LHS  192 &  -143 &  -463 &   -60 &   482  &  LHS 2852 &   -34 &    -5 &    57 &    65  \nl 
  LHS  205a &   119 &  -455 &   -45 &   466  &  LHS 3061 &    73 &  -277 &   103 &   297  \nl 
  LHS  211 &   200 &  -247 &    99 &   307  &  LHS 3073 &  -239 &  -238 &   -76 &   234  \nl 
  LHS  216 &  -321 &  -311 &   -14 &   216  &  LHS 3084 &    31 &  -223 &   -65 &   230  \nl 
  LHS  218 &     4 &  -188 &   -77 &   202  &  LHS 3192 &    45 &  -136 &    72 &   160  \nl 
  LHS  236 &  -207 &  -229 &    93 &   315  &  LHS 3193 &    55 &  -126 &    85 &   160  \nl 
  LHS  276 &  -115 &  -115 &   -47 &   163  &  LHS 3259 &  -189 &  -382 &   125 &   388  \nl 
  LHS  301 &    -7 &  -179 &     8 &   153  &  LHS 3382 &  -302 &  -232 &   -60 &   360  \nl 
  LHS  307 &   173 &  -247 &    33 &   296  &  LHS 3409 &   -30 &  -100 &    42 &    79  \nl 
  LHS  320 &   -38 &  -215 &     0 &   203  &  LHS 3480 &   -13 &   -72 &  -124 &   144  \nl 
  LHS  343 &   107 &  -418 &    89 &   408  &  LHS 3481 &   274 &  -314 &    -6 &   410  \nl 
  LHS  364 &   120 &   -53 &    19 &   132  &  LHS 3548 &   335 &  -331 &   -67 &   470  \nl 
  LHS  375 &    24 &  -188 &   155 &   166  &  LHS 3628 &   273 &  -215 &   112 &   365  \nl 
  LHS  376 &  -173 &  -295 &    98 &   356  &  LHS 3684 &   162 &  -197 &   -58 &   251  \nl 
  LHS  377 &   207 &  -125 &   125 &   205  &  LHS 3867 &  -149 &  -186 &   -78 &   227  \nl 
  LHS  407 &  -215 &  -160 &   -24 &   206  &  LHS 3957 &  -148 &  -125 &    40 &   179  \nl 
  LHS  410 &   165 &  -190 &  -112 &   276  &  TVLM 832-42500 &  -120 &   -22 &     8 &    99  \nl 
  LHS  418 &   279 &  -240 &  -161 &   402  &  
\enddata
\end{deluxetable}

\begin{deluxetable}{rrrrrrrr}
\tablecaption{Velocity Statistics}

\tablewidth{0pt}
\tablenum{6}
\label{velstat}
\tablehead{
\colhead{Group} & 
\colhead{N} &
\colhead{$<V>$} & 
\colhead{$\sigma_U$} & 
\colhead{$\sigma_V$} & 
\colhead{$\sigma_W$} 
}
\startdata
dM   & 16 & -134 & 125 & 78 & 71 \nl
sdM  & 26 & -202 & 177 & 100 & 82 \nl
esdM & 19 & -287 & 176 & 95 & 126 \nl
sdK  & 5 & -238 & 111 & 118 & 61 \nl
strong CaH1 & 24 & -38 & 61 & 44 & 30 \nl
weak CaH1 & 157 & -26 & 48 & 30 & 24 \nl
Disk (RHG) & 514 & -22 & 43 & 31 & 25 \nl
\enddata
\end{deluxetable}

\clearpage


\clearpage

\begin{figure}
\figurenum{1b}
\plotone{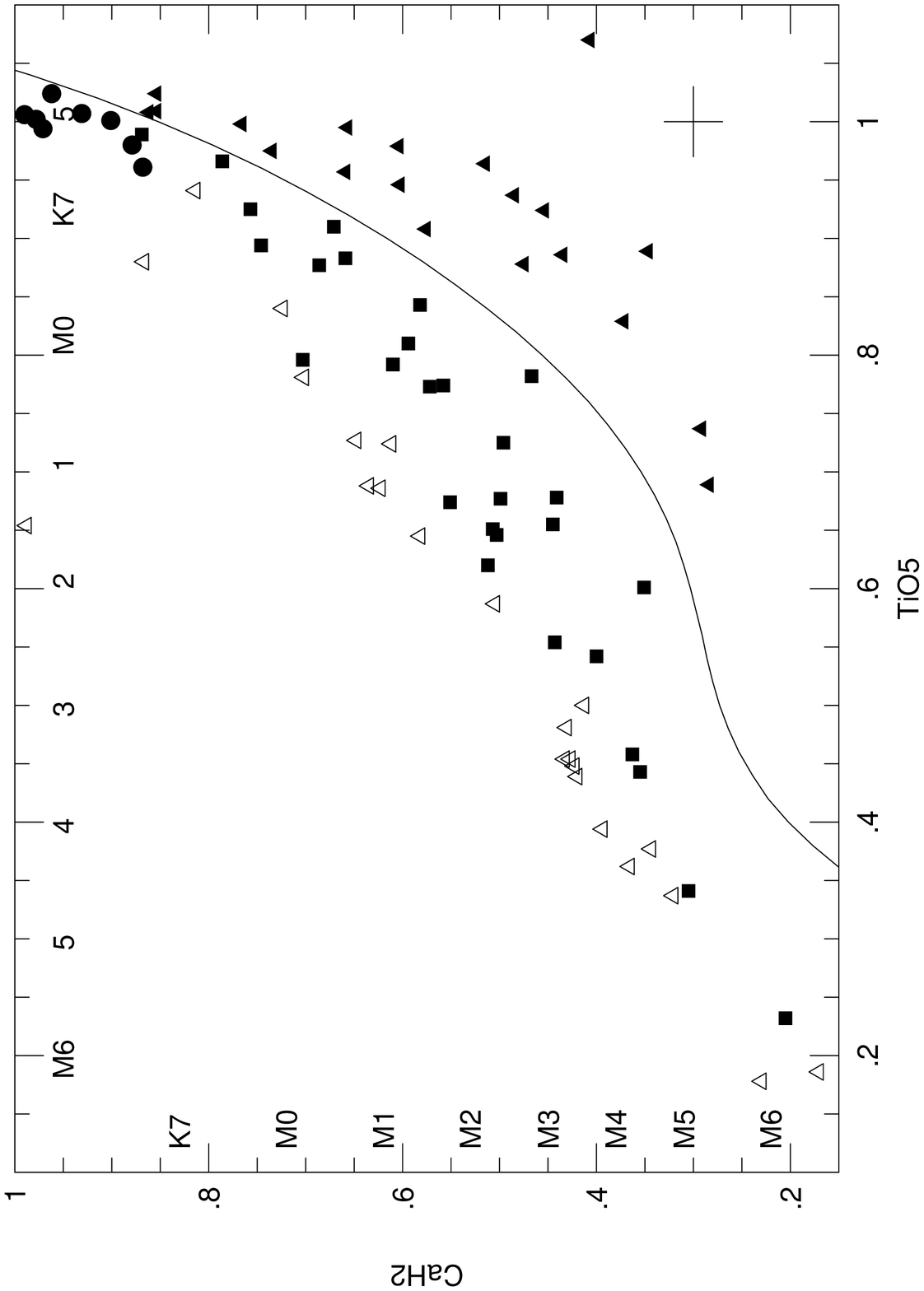}
\end{figure}

\begin{figure}
\figurenum{1c}
\plotone{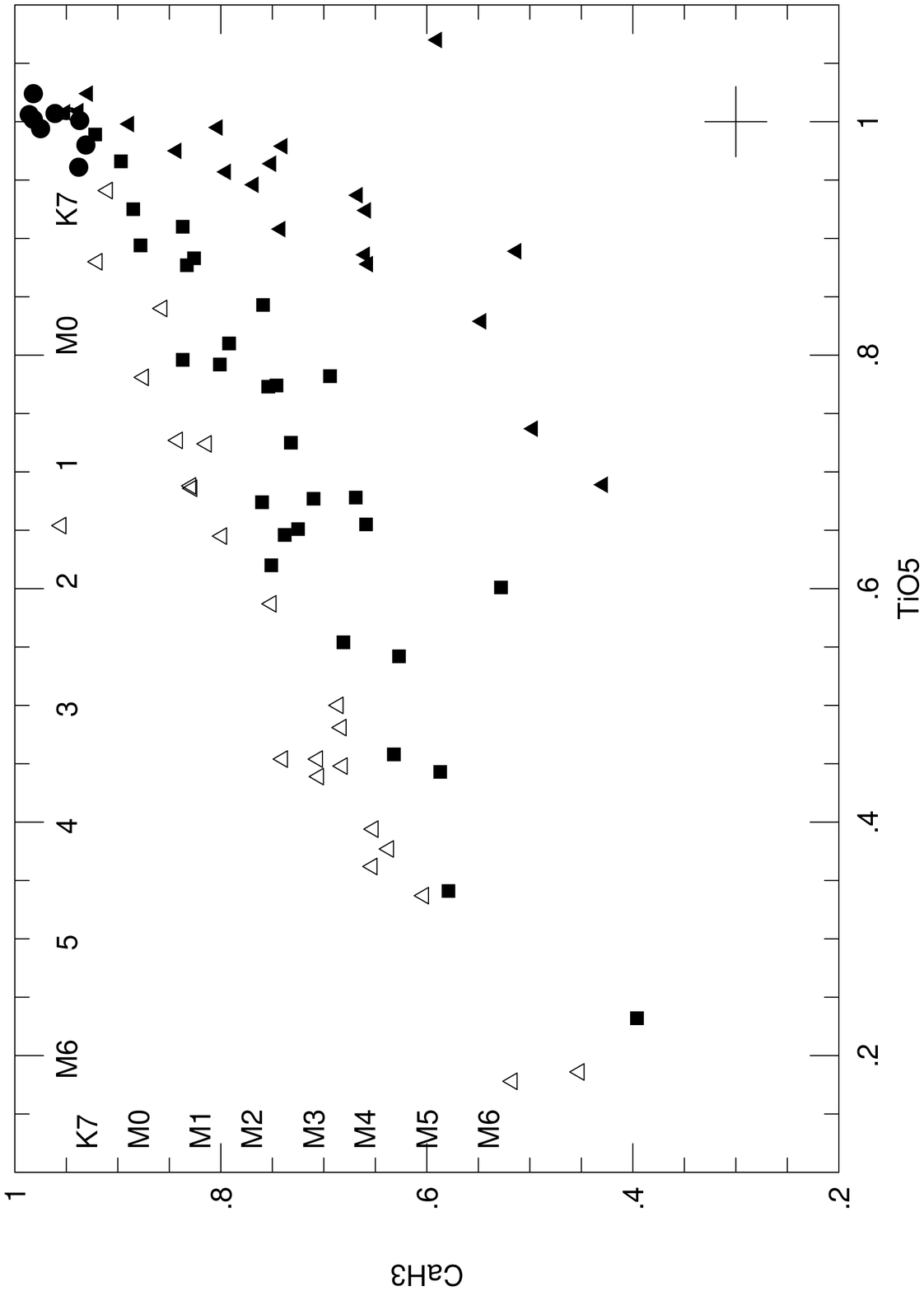}
\end{figure}

\begin{figure}
\figurenum{2}
\plotone{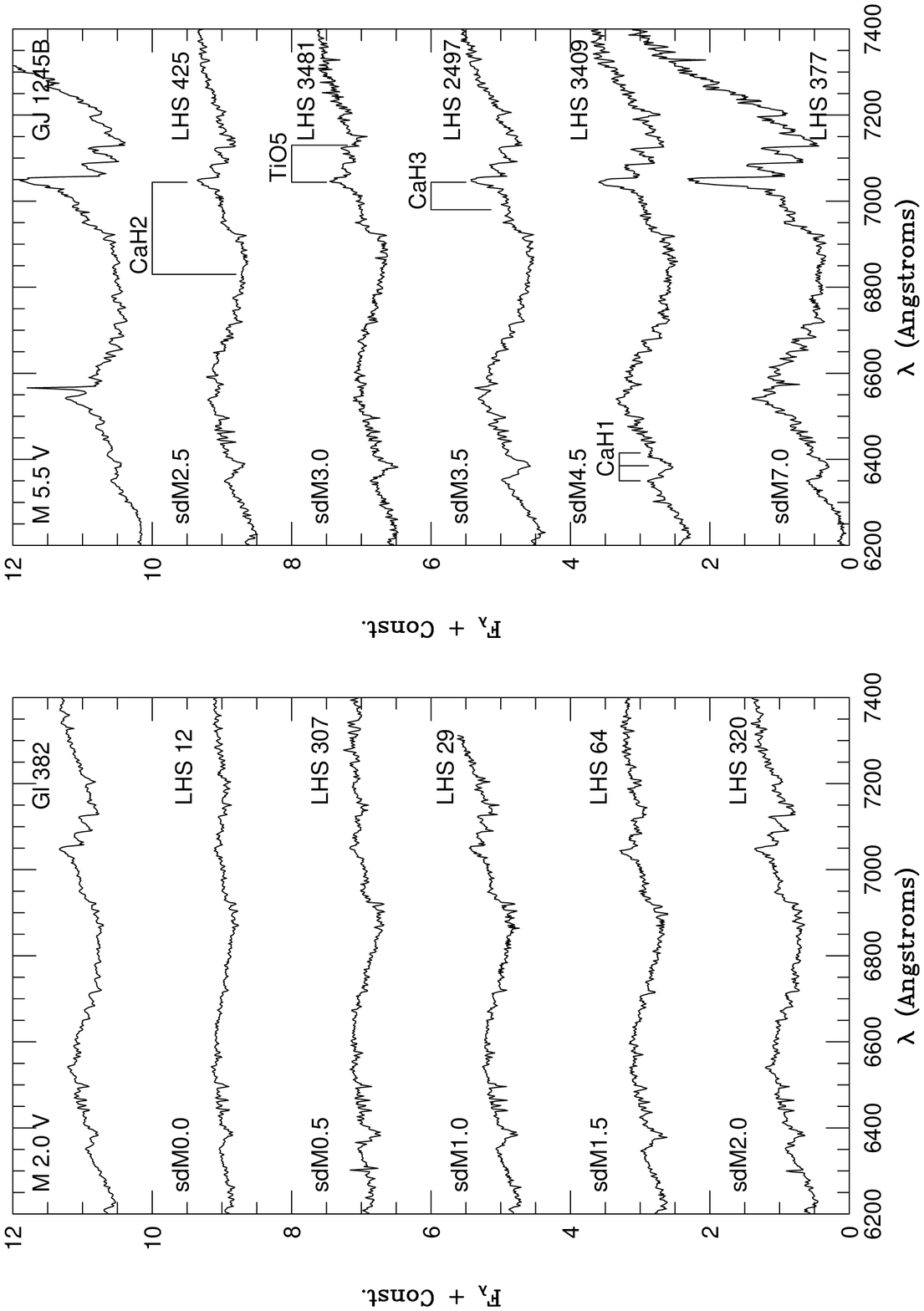}
\caption{A sequence of M-subdwarfs (sdM).  The positions of the 
bandstrength indexes defined in Table 1 appear on the right. 
Two near-solar metallicity 
KHM spectral standards are also shown at the top.\label{figsdm}}
\end{figure}

\begin{figure}
\figurenum{3}
\plotone{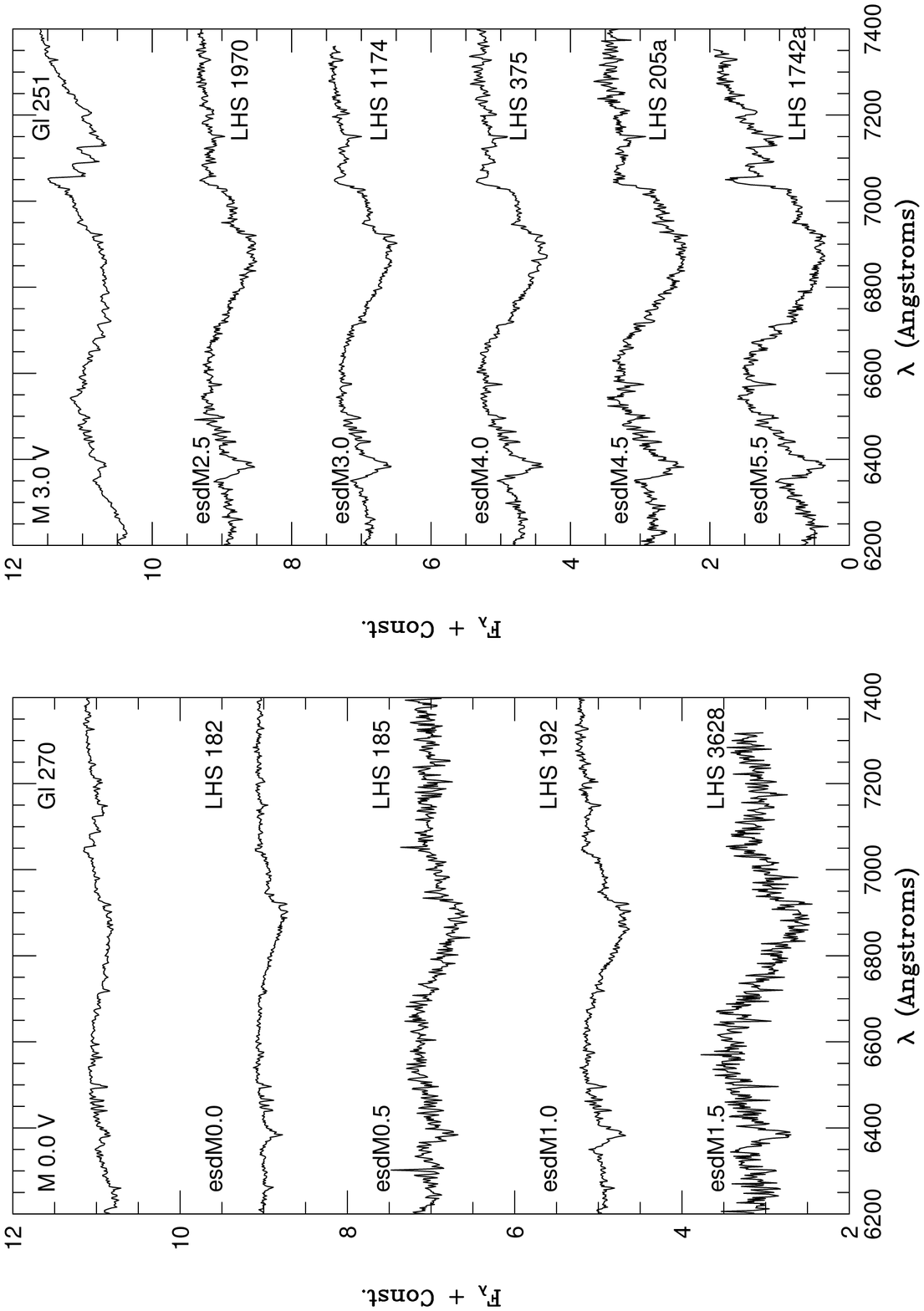}
\caption{A sequence of extreme M-subdwarfs (esdM).  Two near-solar
metallicity KHM spectral standards are also shown at the top. \label{figesdm}}
\end{figure}

\begin{figure}
\figurenum{4}
\plotone{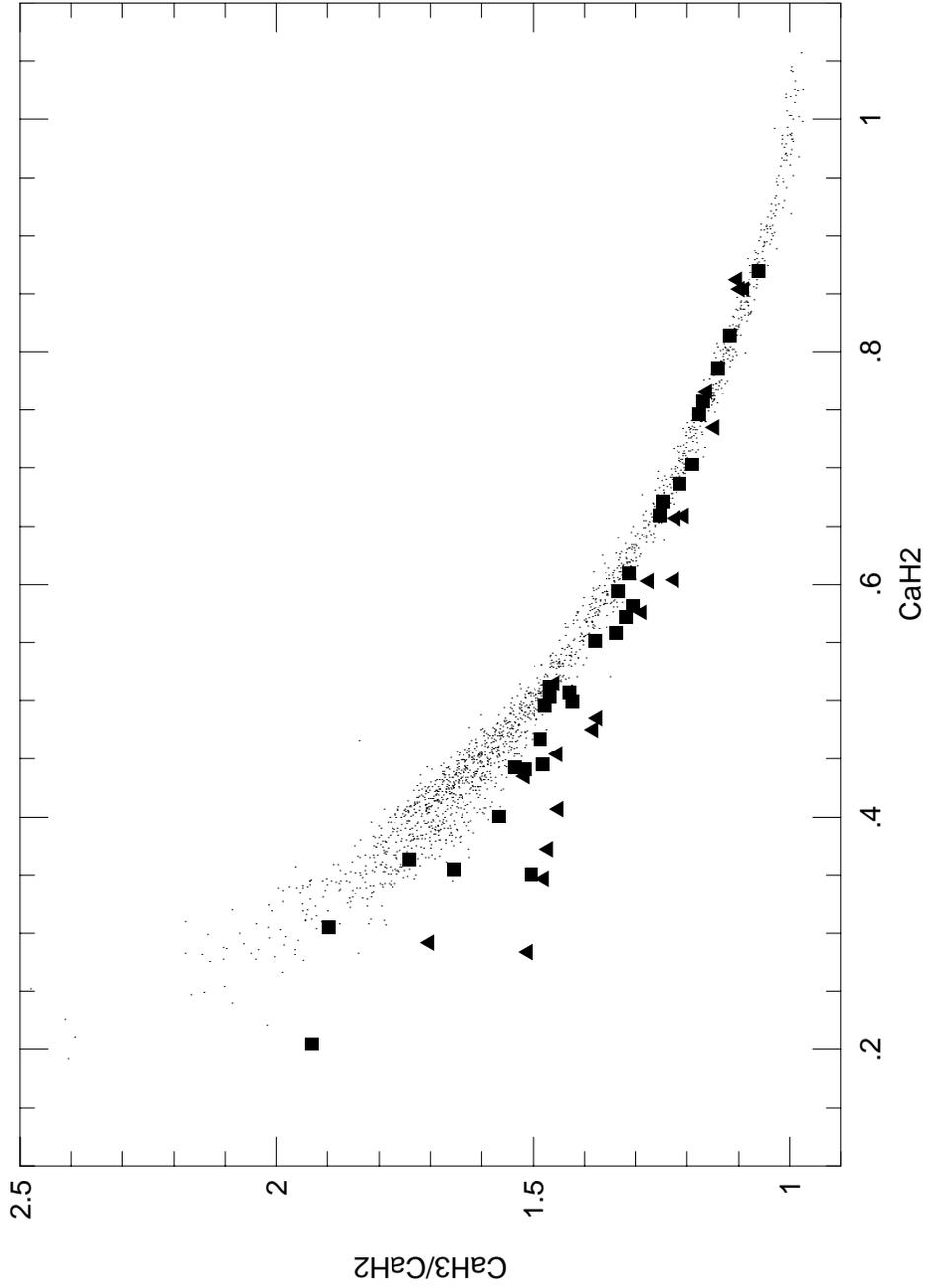}
\caption{The CaH3/CaH2 ratio as function of CaH2.  The 
sdM and esdM lie below the RHG Population I relation. The RHG stars
appear as dots.  \label{ratio}}
\end{figure}

\begin{figure}
\figurenum{5}
\plotone{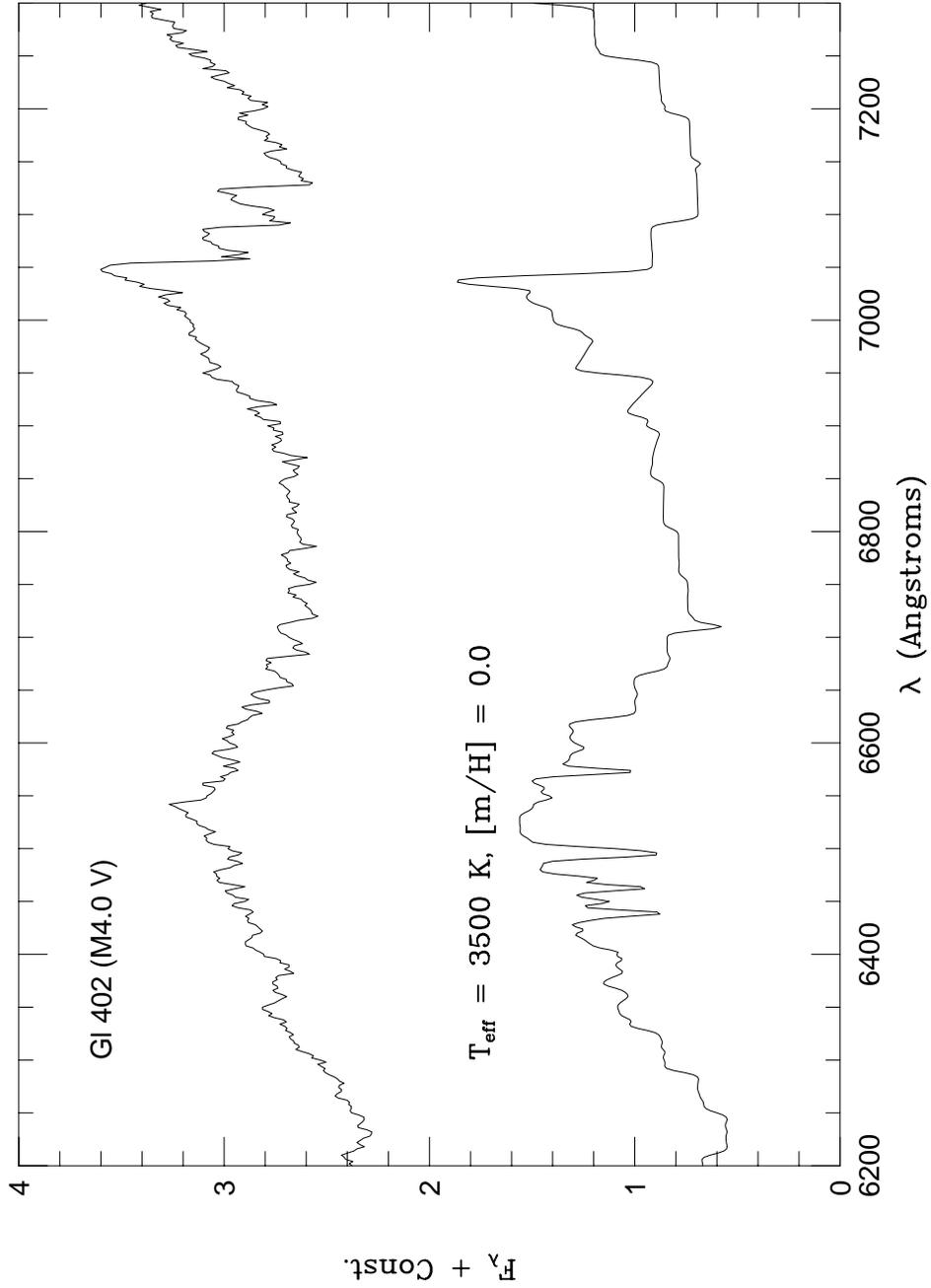}
\caption{A model atmosphere fit to the Population I star Gl 402, 
an M 4.0 V standard. Note that the overly strong atomic lines are
an artifact of the computation technique.  \label{fitpopi}}
\end{figure}

\begin{figure}
\figurenum{6}
\plotone{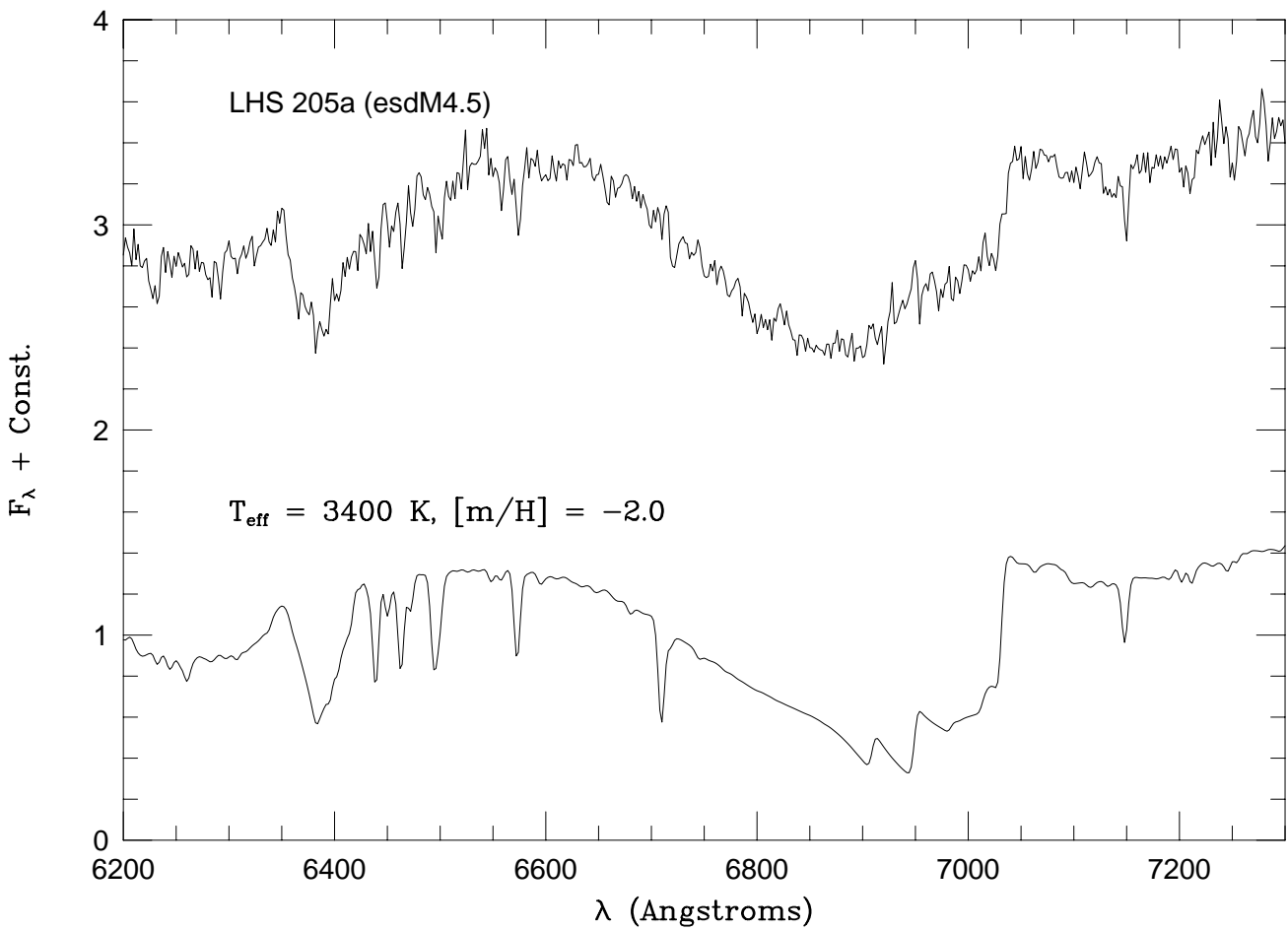}
\caption{Model Atmosphere fit to the extreme subdwarf LHS 205a \label{fitlhs205a}}
\end{figure}

\begin{figure}
\figurenum{7}
\plotone{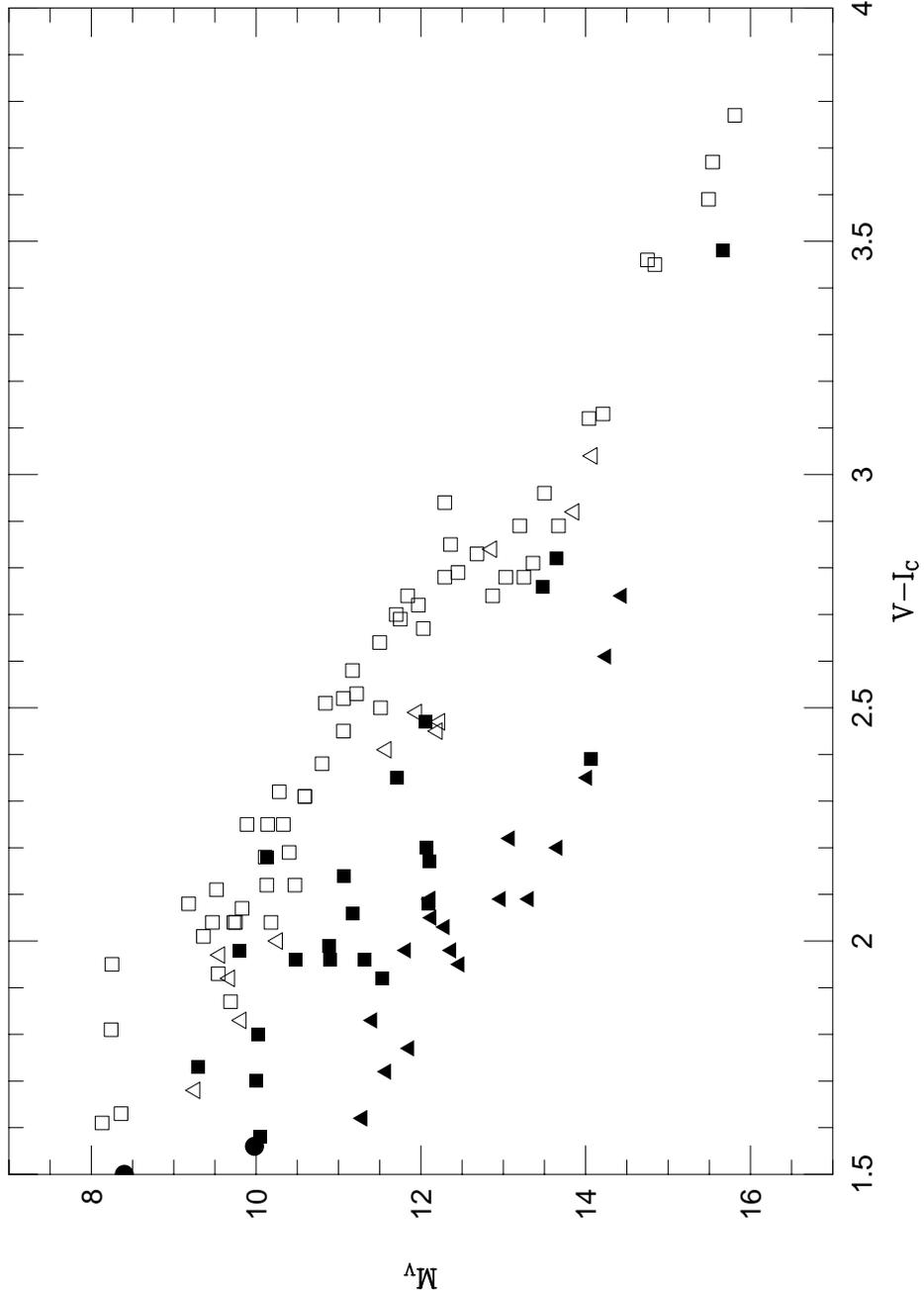}
\caption{The observed HR diagram.  The spectroscopic extreme subdwarfs 
(esdM) are filled triangles, the intermediate subdwarfs (sdM) are filled 
squares, and the subdwarf candidates that were spectroscopically 
indistinguishable are open triangles.
The open squares are the single stars within eight parsecs
with good parallaxes -- note the step at V-I $\sim 2.8$ which is
discussed in HGR. \label{hr}}
\end{figure}

\begin{figure}
\figurenum{8}
\plotone{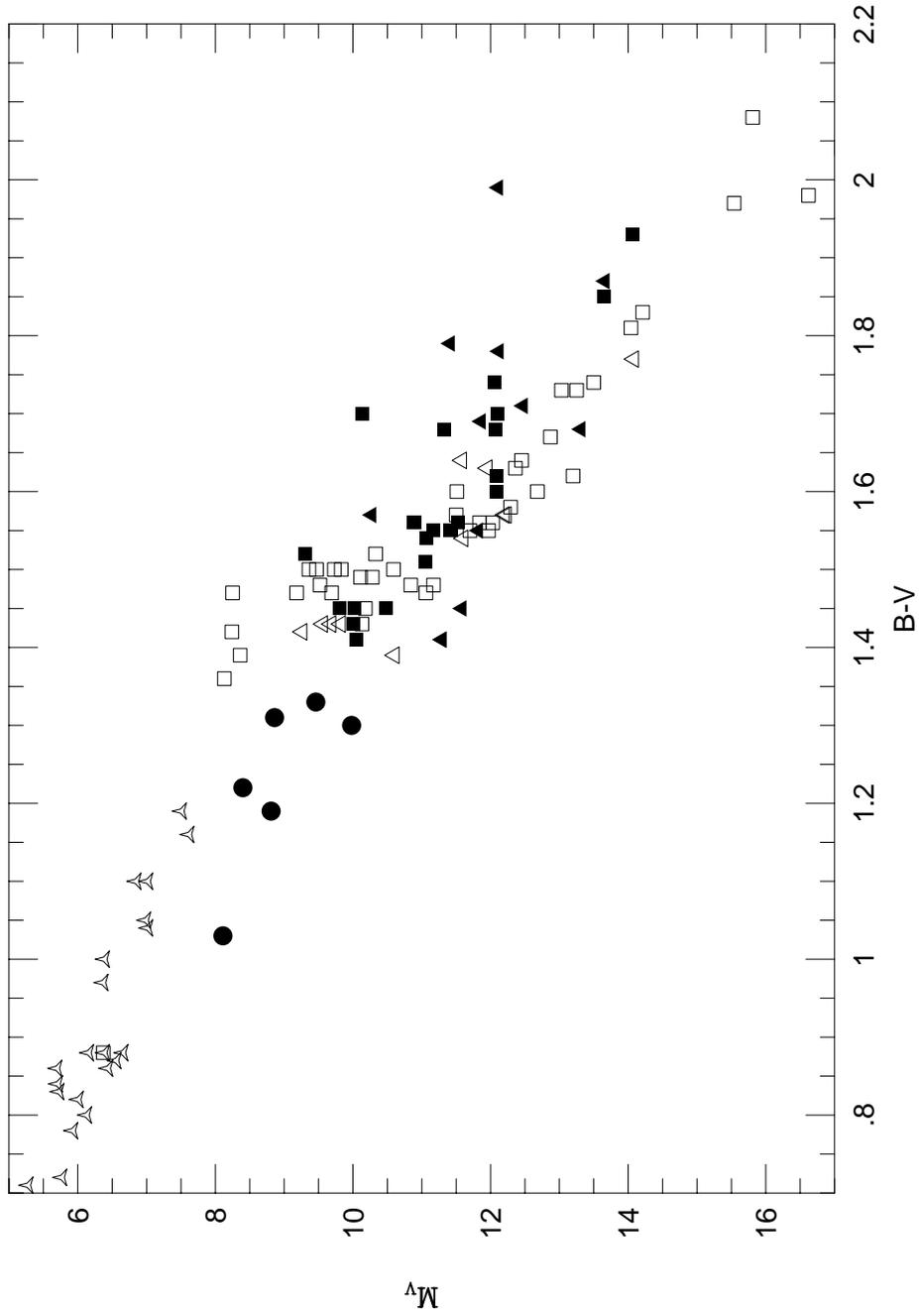}
\caption{The HR diagram for B-V colors.   The three pointed stars
are nearby G and K dwarfs.  Note that the blue K-subdwarfs lie below
the disk sequence, but the sdM and esdM sequences cross the disk
main sequence and are actually brighter at the reddest colors. \label{hrbv}} 
\end{figure}

\begin{figure}
\figurenum{9}
\plotone{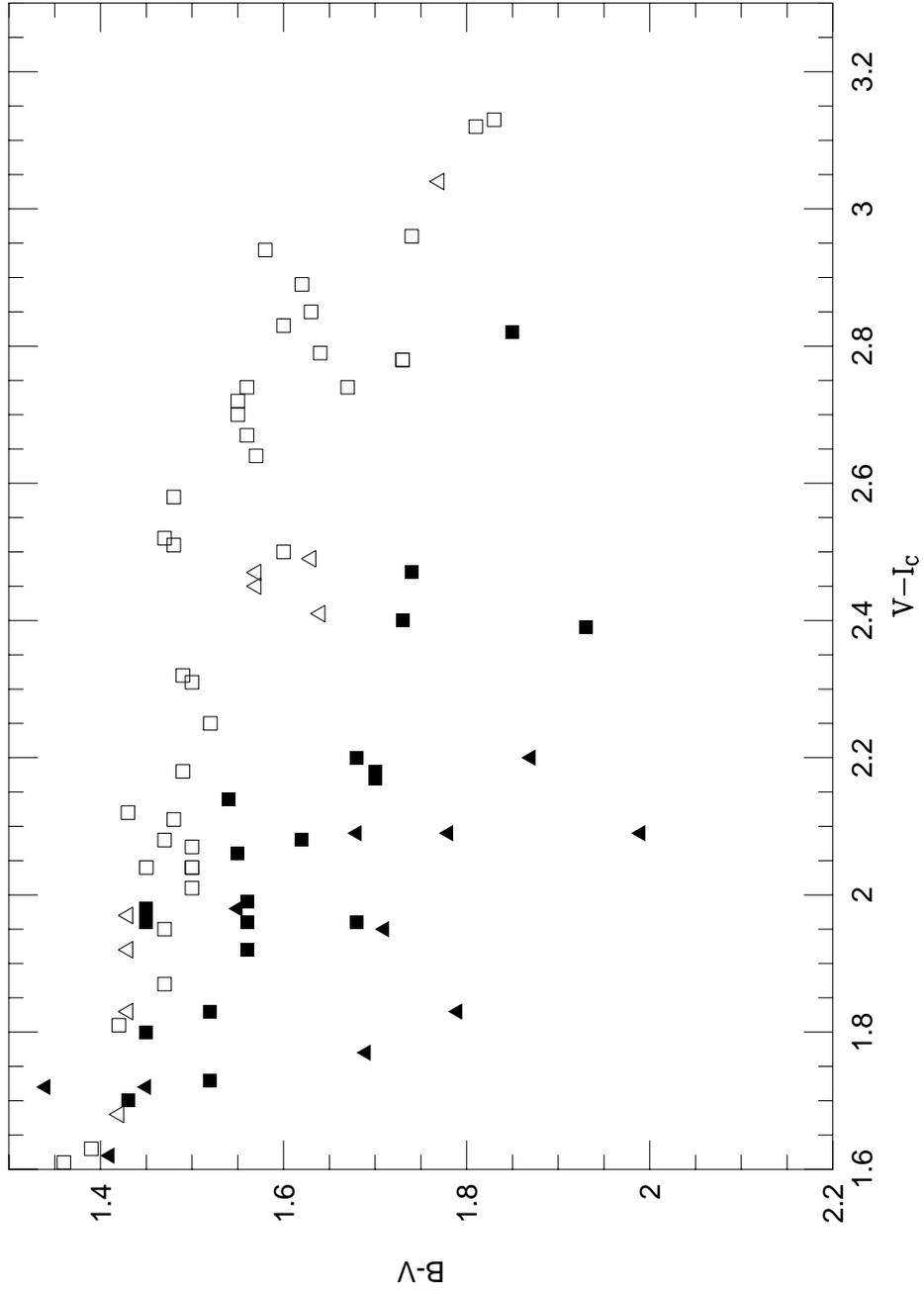}
\caption{V-I vs. B-V.  The sdM are redder in B-V at a given V-I color
than the disk stars.  The esdM show a larger offset than the sdM 
\label{vi-bv}}
\end{figure}

\begin{figure}
\figurenum{10}
\plotone{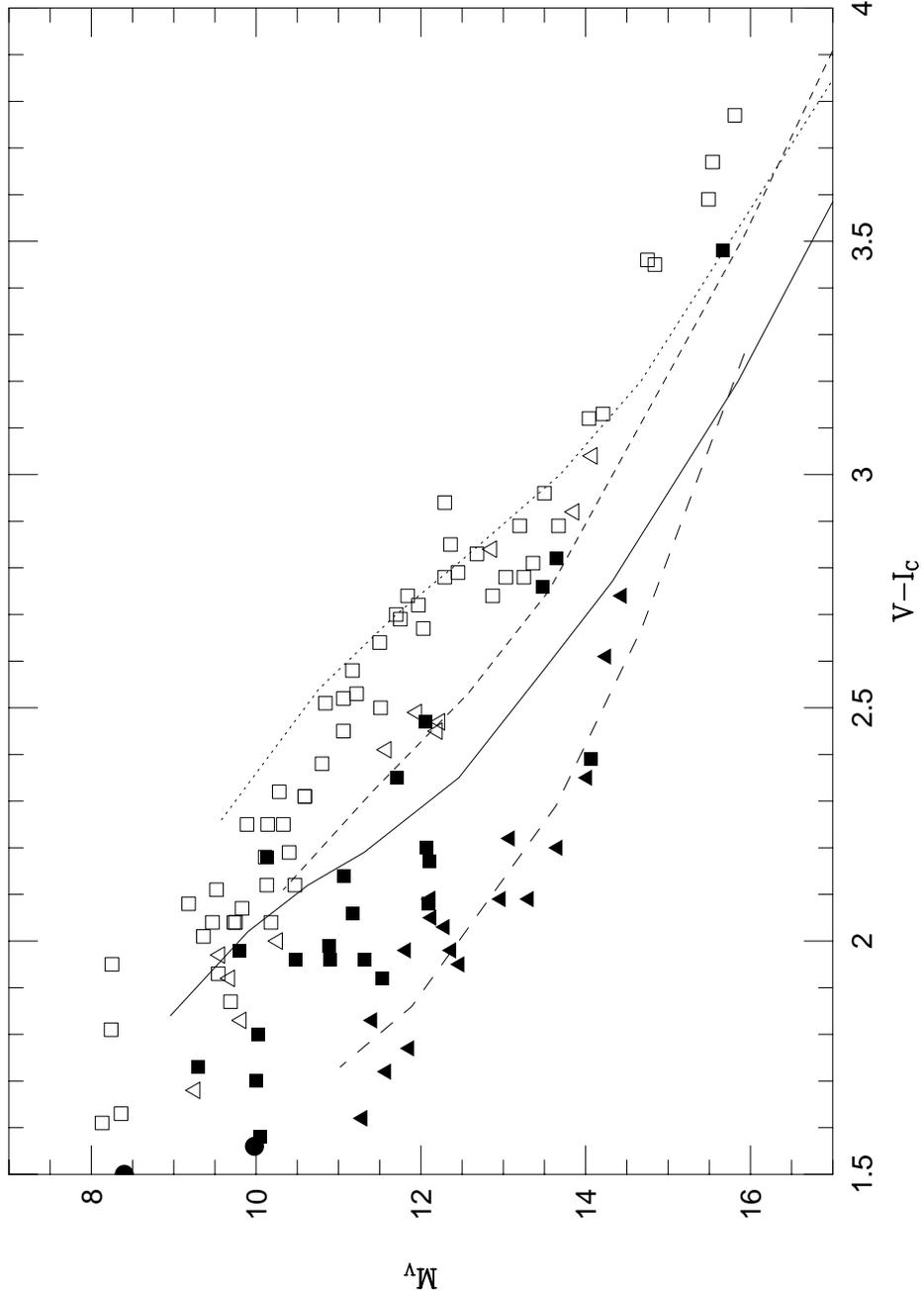}
\caption{The observed HR diagram with theoretical sequences.  
Metallicities of $[m/H] = 0.0$ (dotted line ), -0.5 (short dashed line),
and -1.5 (long dashed line) from Baraffe { et al.} (1995).  A more
recent solar metallicity model (solid line) computed by Baraffe and
Chabrier using the NextGen model atmospheres is also shown.  The latter
model does not match the observed disk stars, but the agreement
of the previous models is quite good.  \label{hrtheory} }
\end{figure}

\begin{figure}
\figurenum{11}
\plotfiddle{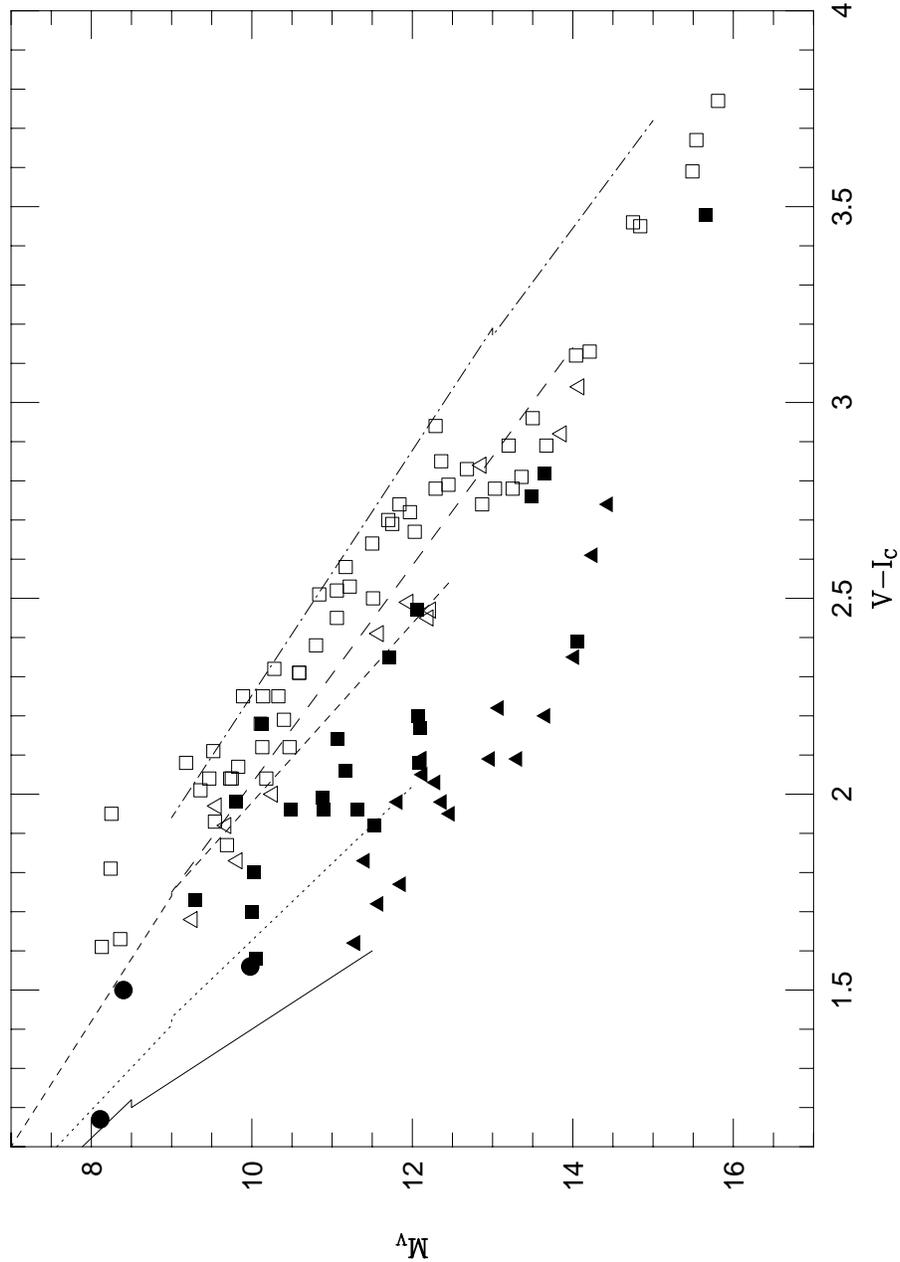}{7in}{90}{70}{70}{216}{0}
\caption{The observed HR diagram with the cluster main sequences
fit by Santiago { et al.}(1996) using HST observations.  
From left to right, the clusters are
M15 (solid line, $[Fe/H] = -2.26$), $\omega$ Cen (dotted, -1.6),
47 Tuc (short dashed, -0.6), N2420 (long dashed, -0.45), and
N 2477 (long dash-dotted, 0.00).  Note that the position of 
the cluster sequences agrees with our spectroscopic metallicity estimates.
The slopes of the cluster sequences do not disagree with the 
local subdwarfs in the region of color overlap when using our
classifications.   The discrepancy of the reddest stars in the
metal rich clusters may be due to color terms, as well as the 
inadequacy of a linear fit for the main sequence.    
\label{hrhst}}
\end{figure}

\end{document}